\documentclass[%
reprint,
superscriptaddress,
amsmath,amssymb,
aps,
prb,
]{revtex4-2}

\usepackage[utf8]{inputenc}
\usepackage{xcolor}
\usepackage{upgreek}
\usepackage{graphicx}
\usepackage{dcolumn}
\usepackage{bm}
\usepackage{natbib}
\usepackage{floatrow}
\usepackage{changes}
\usepackage{eucal}

\DeclareMathAlphabet{\mathcal}{OMS}{cmsy}{m}{n}

\begin{document}
\title{Edge-asymmetric magnetism in semi-Dirac quantum dots}

\author{Mustafa Polat}
\email{mpolat.mustafa@gmail.com}
\affiliation{Medical Imaging Techniques Program, Istanbul Health and Social Sciences Vocational School,\\
Fatih 34122, \.Istanbul, Türkiye}%


\begin{abstract}
In this study, we investigated the electronic and magnetic properties of quantum dots (QDs) throughout the transition from the Dirac limit to the semi-Dirac limit and beyond. When non-interacting semi-Dirac fermions are confined in triangular QDs, the resulting electronic states near the Fermi level are localized asymmetrically among the three edges of these systems. In the presence of electron interactions, we show that these low-energy states favor the formation of edge-asymmetric magnetism in triangular QDs. In particular, the magnetic moments were localized only along a single edge of the zigzag triangular QDs, corresponding to an edge-asymmetric ferromagnetic ground state. For various system sizes, these zigzag triangular QDs exhibit a one-unit increase in the ground-state total spin ($\Delta S_z = +1$) beyond the semi-Dirac limit, revealing a striking result of this transition. In contrast, the remaining two edges are antiferromagnetically coupled in the armchair triangular QDs, indicating an edge-asymmetric antiferromagnetic (AFM) ground state. We further identified the AFM ground states in hexagonal QDs, indicating that these types of asymmetric configurations are not unique to triangular QDs. Collectively, edge-asymmetric magnetism could create potential opportunities for future spintronics.
\end{abstract}

\maketitle

\section{\label{sec:level1}INTRODUCTION}
Since the discovery of graphene \cite{novoselov2004electric,geim2007rise,castro2009electronic}, a broad family of two-dimensional (2D) materials, including hexagonal boron nitride (h-BN) \cite{watanabe2004direct,dean2010boron,cassabois2016hexagonal}, transition metal dichalcogenides (TMDs) \cite{mak2010atomically,splendiani2010emerging,wang2012electronics,chhowalla2013chemistry,manzeli20172d}, Xenes (e.g., silicene, germanene, and stanene) \cite{cahangirov2009two,vogt2012silicene,xu2013large,molle2017buckled}, group-V elemental monolayers (e.g., phosphorene) \cite{liu2014phosphorene,li2014black,carvalho2016phosphorene}, and others \cite{butler2013progress,miro2014atlas,bhimanapati2015recent,khan2020recent}, have been reported over the last two decades. These 2D materials have attracted significant research interest owing to their extraordinary properties and potential applications in next-generation devices \cite{Schaibley2016Valleytronics,burch2018magnetism,
choudhuri2019recent,liu_naturenano_2020,vincent2021opportunities,ozbal2023peculiar}. Furthermore, they provide fertile ground for exploring exotic quantum phenomena such as topological phases \cite{Gao2018Topological,Tokura2019Magnetic,Cayssol2020Topological,Bernevig2022Progress} and magnetic instabilities \cite{meng2022hole}.

Of all the 2D materials that have been reported so far, semi-Dirac systems represent an exceptional class of materials that exhibit an anisotropic dispersion relation, linear in one momentum direction, and quadratic in the perpendicular direction \cite{banerjee2009tight,delplace2010semi,adroguer2016diffusion}. The concept of semi-Dirac fermions emerges from quantum confinement in TiO$_2$/VO$_2$ multilayer heterostructures \cite{Pardo2009} and is realized in a variety of materials and artificial platforms such as $\alpha$-(BEDT-TTF)$_2$I$_3$ organic salt \cite{katayama2006pressure,tajima2009experimental,monteverde2013coexistence}, single-layer black phosphorus \cite{Rodin2014Strain}, topological nodal-line semimetal ZrSiS \cite{Shao2024SemiDirac}, hexagonal close-packed cadmium \cite{Subedi2024DeepLying}, topological insulators \cite{Zhai2011Energy,Li2011Anisotropic,Huang2015Chern}, and photonic lattices \cite{Wu2014SemiDirac,Milicevic2019TypeIII,Real2020Semi-Dirac}. The semi-Dirac dispersion results in a density of states proportional to $D(E)\propto\sqrt{|E|}$ \cite{banerjee2009tight}, a Landau-level spectrum obeying the unconventional relation $E_n(B)\propto[(n+1/2)B]^{2/3}$ \cite{dietl2008new}, and a highly anisotropic interband optical conductivity corresponding to $\sigma_{xx}\propto\sqrt{\Omega}$ and $\sigma_{yy}\propto1/\sqrt{\Omega}$ \cite{carbotte2019optical}, where $\Omega$ is the photon energy. Beyond these properties, the 2D semi-Dirac materials have been extensively examined from various perspectives \cite{Bahamondes2024Holographic,asafov2024collapse,Liao2025Berry,elsayed2025microscopic,ucho2026semidirac}, including their transport properties \cite{link2018out,ye2018semidirac,chan2023quantum,huang2024generation} and topological phase
transitions \cite{Saha2016Photoinduced,garcia2026corner,xu2026floquet}.

\begin{figure}[b]
	\begin{center}
		\includegraphics[width=1\linewidth, height=\textheight,keepaspectratio]{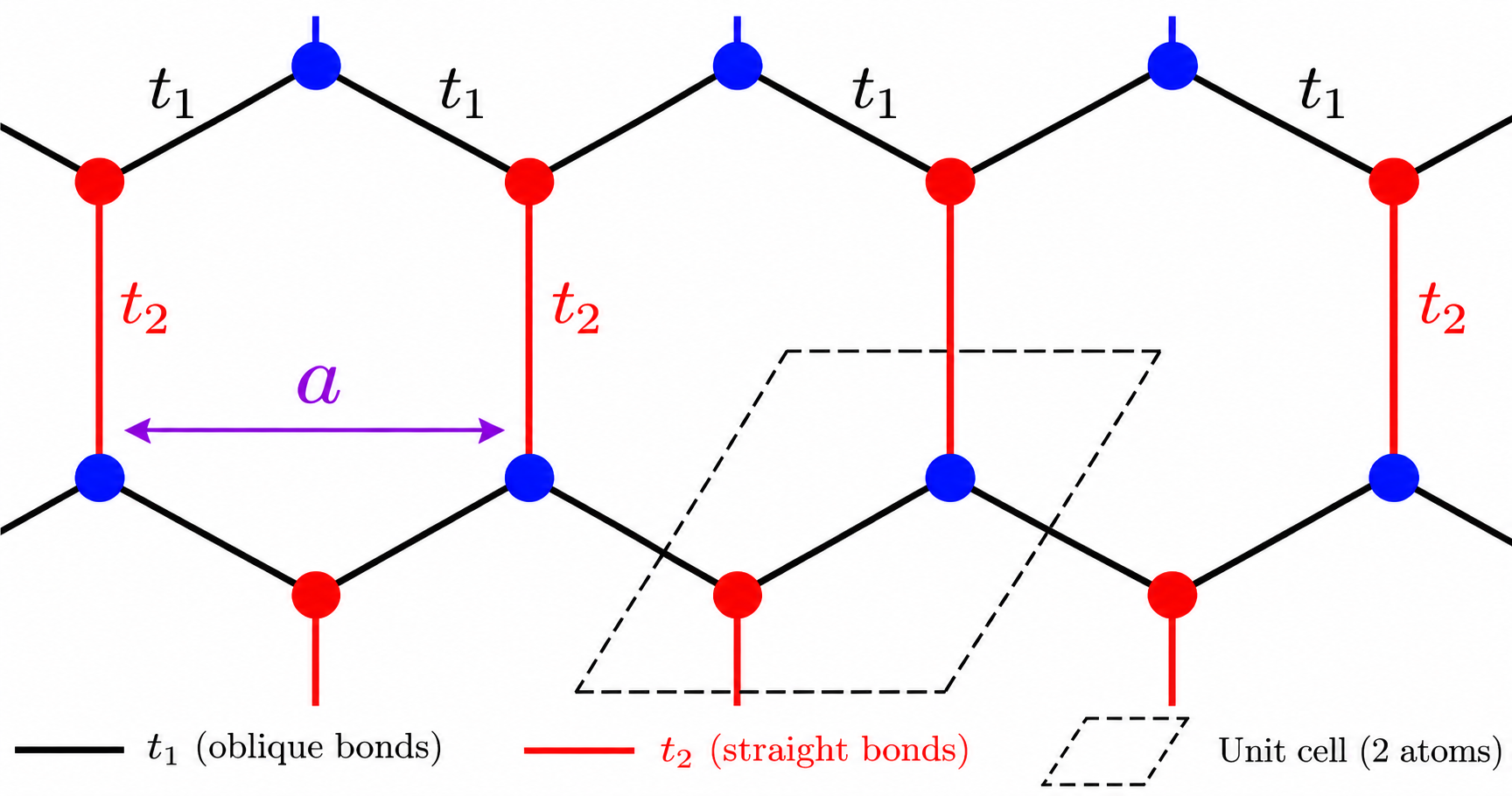}
		\caption{Schematic representation of the TB model on a honeycomb lattice, involving the nearest-neighbor hopping amplitudes $t_1$ and $t_2$ along the oblique and straight bonds, respectively. The lattice constant is set to $a = 1$, and the unit cell containing A (red) and B (blue) sublattices is shown with the dashed parallelogram.}
\label{fig_1} 
	\end{center}
\end{figure}

\begin{figure*}
	\floatbox[{\capbeside\thisfloatsetup{capbesideposition={right,center},capbesidewidth=6.21cm}}]{figure}[\FBwidth]
	{\caption{Evolution of two Dirac points around the $M$ point as a function of hopping anisotropy $t_2/t_1$ within the TB model. The merging of two Dirac points occurs at the $t_2/t_1=2$, signaling an anisotropic dispersion relation, linear in one momentum direction and quadratic in the perpendicular direction. For $t_2/t_1 > 2$, the insulating regime appears in this 2D system.}\label{fig_2}}
	{\includegraphics[width=4.4in]{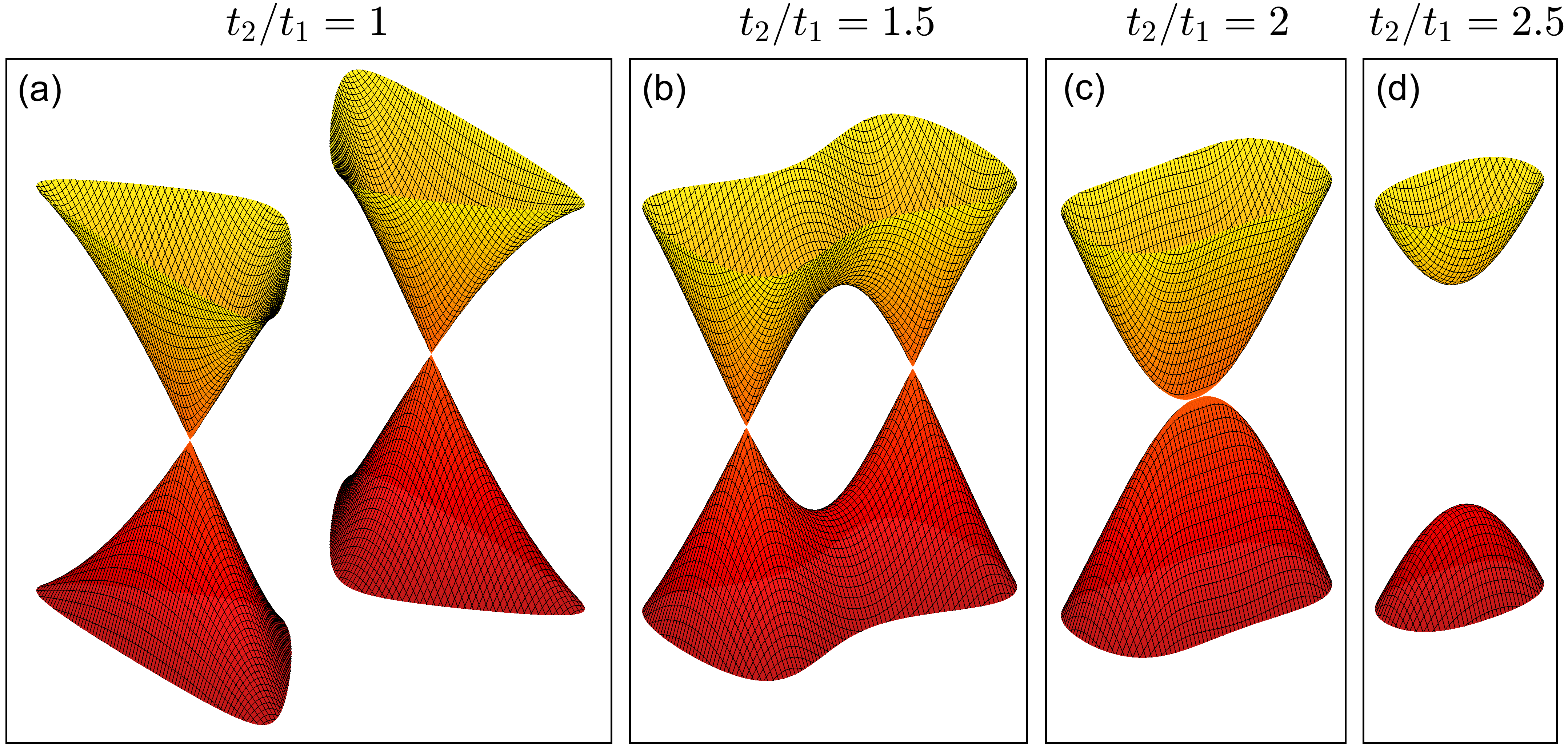}}
\end{figure*}

In recent years, the semi-Dirac systems of finite width have attracted special interest \cite{GarcíaOlmos2025Spin-dependent}, and it has been reported using a $k\cdot p$ Hamiltonian that the energy-dependent edge states are localized on either the upper or lower edge of a nanoribbon (NR) depending on their electron–hole character \cite{GarcíaOlmos2024Zero}. In another study, anisotropic magneto-transport behavior was predicted in zigzag semi-Dirac NRs \cite{sinha2020quantum} using the tight-binding (TB) model on a honeycomb lattice \cite{hasegawa2006zero,montambaux2009merging,montambaux2009universal,hasegawa2012merging,chen2018floquet}, as depicted in Fig.~\ref{fig_1}. These recent findings reveal the significant consequences of reduced dimensionality in semi-Dirac physics; however, the interplay between complete quantum confinement and hopping anisotropy remains largely unexplored in the existing literature. 

In this study, we create triangular and hexagonal quantum dots (QDs) of various sizes with zigzag and armchair edge terminations and systematically investigate their electronic and magnetic properties throughout the transition from the Dirac limit ($t_2/t_1 = 1$) to the semi-Dirac limit ($t_2/t_1 = 2$) and beyond ($t_2/t_1 > 2$). In the non-interacting regime, the relative localization of electronic states along the three edges of triangular QDs is quantitatively analyzed through the energy-resolved pairwise edge anisotropies, and the results indicate that the anisotropy-induced spatial redistribution of low-energy states triggers the formation of edge-asymmetric ground states in these zero-dimensional (0D) systems. For all system sizes considered here, the self-consistent mean-field Hubbard (MFH) calculations show that the armchair and zigzag triangular QDs develop edge-asymmetric antiferromagnetic (AFM) and ferromagnetic (FM) ground states, respectively. In particular, the magnetic moments are localized only along the bottom edge of the zigzag triangular QDs, whereas the remaining two edges are antiferromagnetically coupled in the case of armchair triangular QDs. We also show that the ground-state total spin of the zigzag triangular QDs exhibits an increase of $\Delta S_z = +1$ when $t_2/t_1 > 2$. This is due to the genuine level crossing, which is clearly evident in the discrete energy spectrum. Furthermore, we identify the edge-asymmetric AFM ground states in hexagonal QDs in the semi-Dirac limit. For all system geometries, the asymmetric behavior at the edges becomes more pronounced near the semi-Dirac limit and beyond. Overall, the results presented in this paper may pave the way for understanding the mechanism of edge-asymmetric magnetism in the finite-sized structures.

The rest of this paper is organized as follows. In Sec.~\ref{sec:methods}, the computational methods are introduced. The edge-asymmetric magnetization in triangular and hexagonal QDs of various sizes is discussed in Sec.~\ref{sec:results}. Sec.~\ref{sec:conclusions} summarizes our conclusions.
\section{\label{sec:methods} COMPUTATIONAL METHODS}
We mainly study edge magnetism in the QDs through extended MFH model at the $\pi_z$ level, which can successfully capture the electronic and magnetic properties of nanostructures in agreement with the first-principles calculations \cite{fernandez2007magnetism,yazyev2010emergence,li2020uncovering}. It can be written for the interacting electrons on a honeycomb lattice as follows:

\begin{align}
H_{\mathrm{MFH}}
=&
-\sum_{\langle i,j\rangle,\sigma}
t_{ij}
\left(
c_{i\sigma}^{\dagger}c_{j\sigma}
+\mathrm{H.c.}
\right)
\nonumber\\
&
+U\sum_{i,\sigma}
\left(
\langle n_{i\bar{\sigma}}\rangle-\frac{1}{2}
\right)
\hat{n}_{i\sigma}
\nonumber\\
&
+\sum_{i,j}
V_{ij}
\left(
\langle n_j\rangle-1
\right)
\hat{n}_i
\nonumber\\
&
-U\sum_i
\left(
\langle n_{i\uparrow}\rangle-\frac12
\right)
\left(
\langle n_{i\downarrow}\rangle-\frac12
\right)
\nonumber\\
&
-\sum_{i,j}
V_{ij}
\left(
\langle n_i\rangle-1
\right)
\left(
\langle n_j\rangle-1
\right),
\label{MFH}
\end{align}
where $c^{\dagger}_{i{\sigma}}$ ($c_{i{\sigma}}$) is a creation (annihilation) operator for an electron on the site $i$ with spin $\sigma$. In this study, we set the nearest-neighbor hopping amplitude  to $t_{ij} = t_1$ for the oblique bonds and $t_{ij}=t_2$ for the straight bonds, where $t_1 = 1\,\mathrm{eV}$ and $1\,\mathrm{eV}\leq t_2\leq 2.5\,\mathrm{eV}$ (see also Fig.~\ref{fig_1}). 

We fix the on-site Coulomb interaction to $U = 0.5t_1$ for all self-consistent calculations \cite{feldner2010magnetism}. In Eq.~\eqref{MFH}, \(\langle n_{i\sigma}\rangle\) and $\hat{n}_{i\sigma}$ are the spin-resolved site occupation and number operator, respectively. The total occupation is given by \(\langle n_i\rangle\), and $\hat{n}_i$ is the total number operator in the extended MFH model. We take the off-site interaction as \(V_{ij} = V_{\mathrm{off}}/r_{ij}\), where \(r_{ij}=|\mathbf{r}_i-\mathbf{r}_j|\) is the dimensionless distance between two sites, and $V_{\mathrm{off}} = U/2$. The off-site Coulomb interaction is extended to all lattice sites in this study. The final two constant terms in Eq.~\eqref{MFH} are the mean-field energy contributions. The unbiased self-consistent calculations are performed without any constraints regarding the sublattice imbalance, whereas the total number of electrons is always kept at constant by assuming the half- filled case. 

We first apply the TB model to the 2D system to examine the merging of two Dirac points, and this is achieved by systematically tuning the hopping anisotropy $t_2/t_1$. Fig.~\ref{fig_2} (a)-(d) show the low-energy electronic states of 2D system within the energy window $E \in [-0.9,\,0.9]\,\mathrm{eV}$, where the Fermi level (FL) is set zero. The calculations are performed on a $401 \times 401$ $k$-point grid in the first Brillouin zone. The merging of two Dirac points occurs at the $M$ point $\left(0,2\pi/\sqrt{3}\right)$ when $t_2/t_1 = 2$, as displayed in Fig.~\ref{fig_2} (c). As the hopping anisotropy $t_2/t_1$ is further increased, the insulating regime appears in this 2D system at $t_2/t_1 = 2.5$, as shown in Fig.~\ref{fig_2} (d). These results correspond to a transition between a semi-metal and a band insulator in the non-interacting case.

\begin{figure}[b]
	\begin{center}
		\includegraphics[width=1\linewidth, height=\textheight,keepaspectratio]{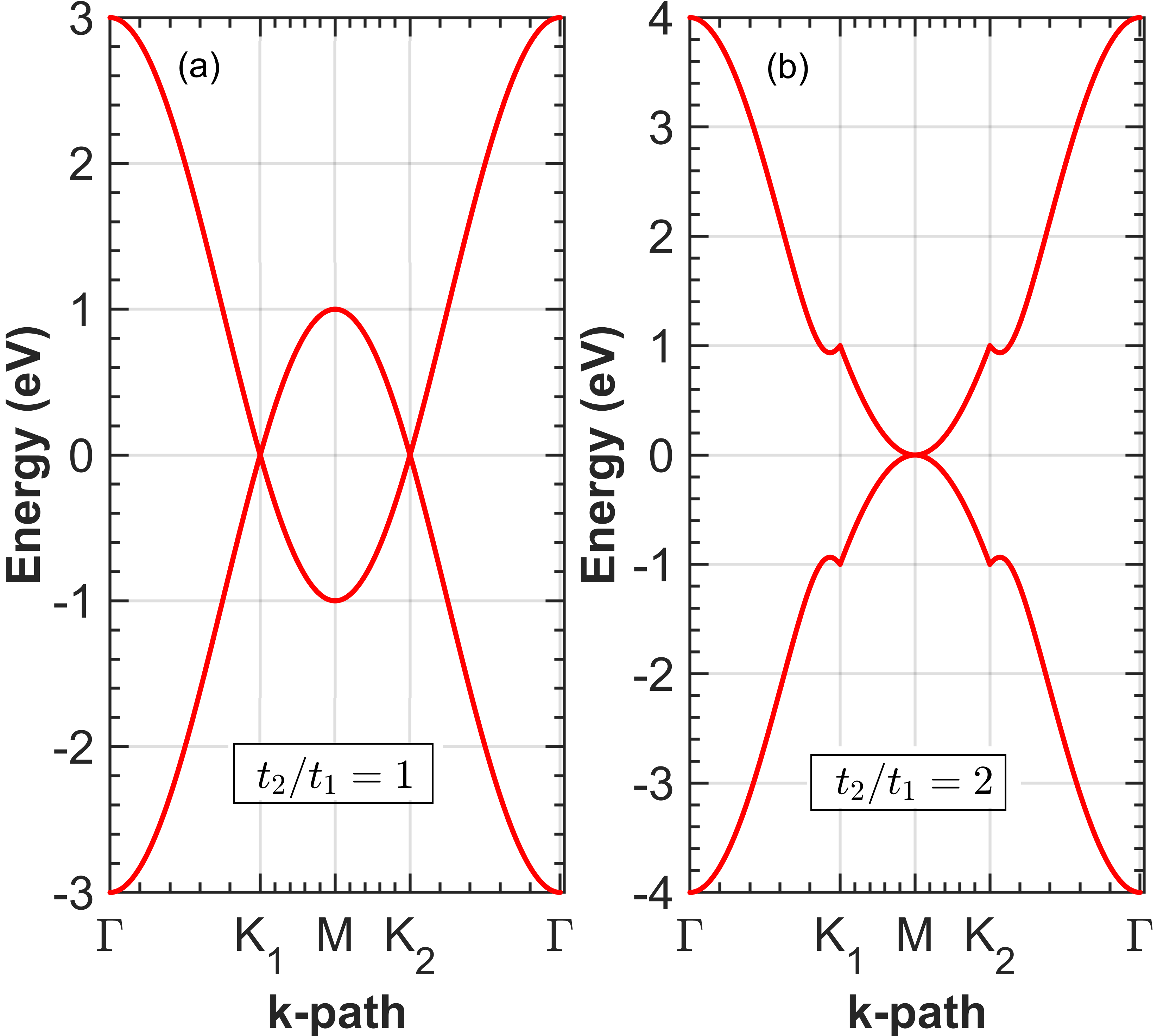}
		\caption{The 2D band structures calculated using the TB model.  These energy spectra along the high symmetry $\Gamma-K_1-M-K_2-\Gamma$ path are shown in (a) the Dirac and (b) semi-Dirac limits.}
\label{fig_3} 
	\end{center}
\end{figure}

The merging of two Dirac points is also studied through 2D band structures within the TB model, as shown in Fig.~\ref{fig_3} (a)-(b). In the left panel, the band structure along the high symmetry $\Gamma-K_1-M-K_2-\Gamma$ path is shown in the Dirac limit. At $E=0\,\mathrm{eV}$, the band dispersion clearly exhibits two band crossings at $K_1$ $\left(2\pi/3,2\pi/\sqrt{3}\right)$ and $K_2$ $\left(-2\pi/3,2\pi/\sqrt{3}\right)$ points, associated with two distinct Dirac points. At the $M$ point, the valance and conduction bands touch $+1~\mathrm{eV}$ and $-1~\mathrm{eV}$, respectively. On the other hand, the two crossings at $K_1$ and $K_2$ points disappear in the semi-Dirac limit, as show in the right panel. At $t_2/t_1 = 2$, the upper and lower bands meet at the $M$ point, indicating the merging of two Dirac points. The energy range also extends from $\pm3~\mathrm{eV}$ in the left panel to $\pm4~\mathrm{eV}$ in the right panel due to an increase in $t_2/t_1$. These results demonstrate that anisotropic nearest-neighbor hopping alone is sufficient to produce a significant  reconstruction of the energy spectrum.

\begin{figure}[t]
	\begin{center}
		\includegraphics[width=\linewidth, height=\textheight,keepaspectratio]{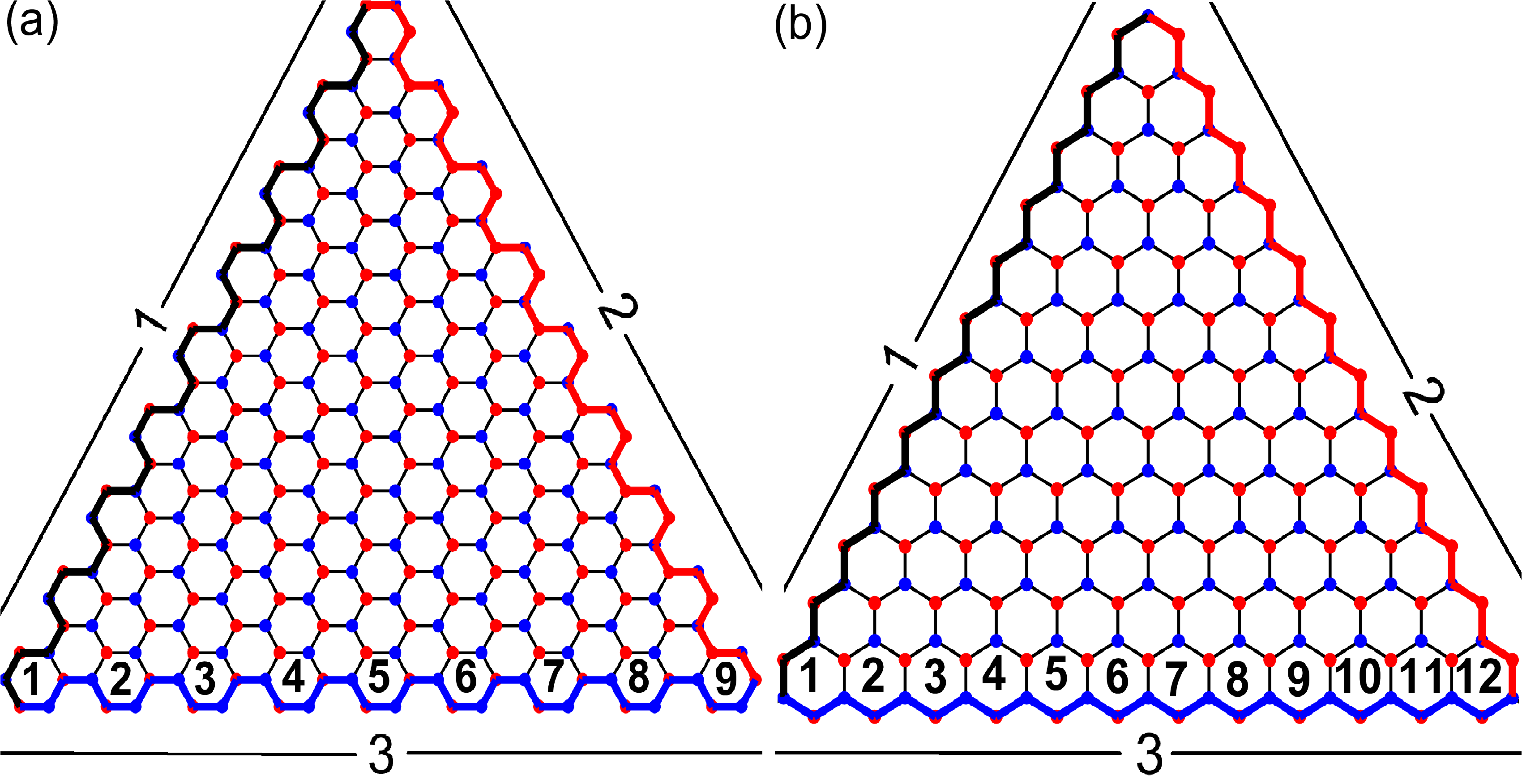}
		\caption{The representative system geometries of (a) armchair and (b) zigzag triangular QDs with $N_{\mathrm{edge}} = 9$ and $12$, respectively. The edge atoms extending along each side of the triangular systems are shown through the black, red, and blue segments.}
\label{fig_4} 
	\end{center}
\end{figure}

We select QDs as a platform to probe the interplay between complete quantum confinement and hopping anisotropy. These QDs have received considerable attention due to their tunable geometric features, such as system size, shape, and edge termination and find applications in a variety of fields, ranging from biomedicine to electronics \cite{yan2019recent}. The representative system geometries of triangular QDs are shown in Fig.~\ref{fig_4} (a)-(b). These QDs are classified with respect to their edge terminations: (a) armchair and (b) zigzag edges. The size of each equilateral triangular QD is given by $N_{\mathrm{edge}}$, which is defined as the number of hexagons along a single edge of the triangle. Here, the corresponding system sizes are $N_{\mathrm{edge}} = 9$ (left panel) and $12$ (right panel), respectively. The edges of these systems are indexed as $1$, $2$, and $3$ to discriminate these three edges from each other.

\begin{figure*}[t]
   \centering
    \includegraphics[width=1\textwidth]{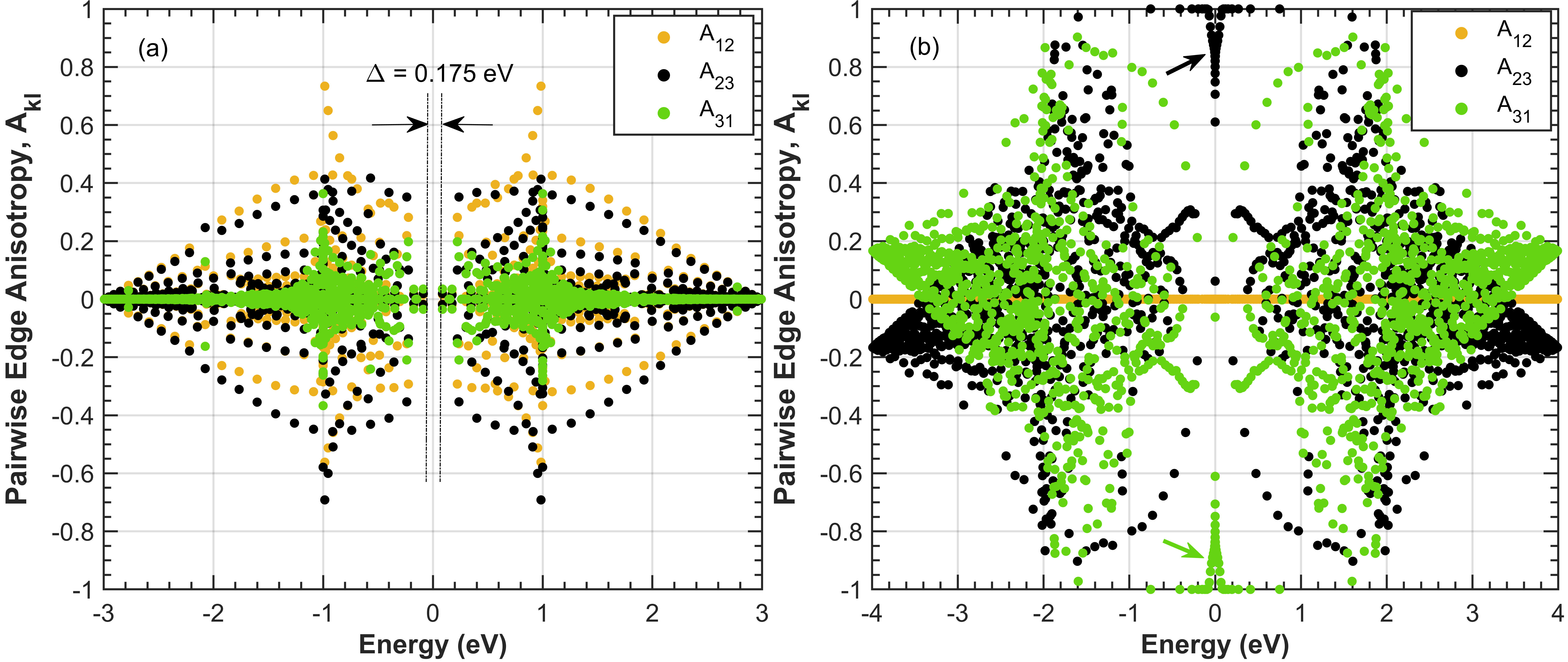}
    \caption{The energy-resolved pairwise edge anisotropies obtained from the TB method. The left and right panels show the results of the armchair triangular QDs with $N_{\mathrm{edge}} = 46$ in the Dirac and semi-Dirac limits, respectively. In the right panel, the accumulations near the FL, marked by the colored arrows, indicate the existence of edge-asymmetric localization of the low-energy states in this system.}
\label{fig_5}
\end{figure*}

The relative localization of electronic states along the three edges of triangular QDs are investigated through the pairwise edge anisotropy $A_{kl} (E_n)$, which is defined as follows:

\begin{equation}
A_{kl}(E_n)=
\frac{w_k(E_n)-w_l(E_n)}
{w_k(E_n)+w_l(E_n)},
\label{PEA}
\end{equation}
where the indices $k$ and $l$ denote the edge indexes. The probability density of the $n$-th eigenstate is summed over the atoms extending along the corresponding edge (see also the colored segments in Fig.~\ref{fig_4}). This can be written as $w_k(E_n) = \sum_{i\in k}|\psi_n(i)|^2$, where $E_n$ is the energy of the $n$-th eigenstate. As is clear from Eq.~\eqref{PEA}, $A_{kl}$ is zero when $k$ = $l$. As for $k \neq l$, $A_{kl}$ approaches unity if the $n$-th eigenstate is primarily localized only along single edge of the system. The energy-resolved pairwise edge anisotropies obtained from the TB method support our mean-field results by exhibiting imbalanced edge localization in triangular QDs. 

The spin variance is addressed for the purpose of measuring spatially inhomogeneous spin polarization in the QDs. The spin variance is defined as follows:

\begin{equation}
\sigma_{S_z}^{2}
=
\frac{1}{N}
\sum_{i=1}^{N}
\left(
S_i-\langle S_z\rangle
\right)^2,
\label{spinvariance}
\end{equation}
where $S_i= 1/2(\langle n_{i\uparrow}\rangle-\langle n_{i\downarrow}\rangle)$ is the local spin polarization at site $i$, N is the number of total atoms, and $\langle S_z\rangle=\sum_{i=1}^{N}S_i/N$ is the average spin polarization of the analyzed system. 
\section{Results and Discussion}\label{sec:results}
We find the lowest-energy solution for each system geometry and present a more detailed explanation of the properties of these ground states on the following pages. 
\subsection{Armchair Triangular QDs}
We first calculate the energy-resolved pairwise edge anisotropies of the armchair triangular QDs with $N_{\mathrm{edge}} = 46$. The TB calculations are separately performed in the Dirac and semi-Dirac limits, and the corresponding results are shown in Fig.~\ref{fig_5} (a)-(b), respectively. In both limits, we only consider the cyclic edge pairs of $(1,2)$, $(2,3)$, and $(3,1)$, yielding $A_{12}(E)$ (orange distribution), $A_{23}(E)$ (black distribution), and $A_{31}(E)$ (green distribution), respectively. 

In the Dirac limit, these three pairwise edge anisotropies remain tightly confined around zero across the entire energy range $-3\,\mathrm{eV}\leq E\leq 3\,\mathrm{eV}$. As shown in Fig.~\ref{fig_5} (a), a large number of data points satisfy the condition, that is, $|A_{kl}(E)| < 0.4$. Despite being confined to a narrow range of values, these distributions are highly symmetrical with respect to $E = 0\,\mathrm{eV}$. This means that the electron-hole symmetry is preserved in this limit. In addition, all distributions vanish in the vicinity of the FL, resulting in a band gap of $\Delta = 0.175\,\mathrm{eV}$. These pairwise edge anisotropies form flat bands near the FL, implying equal edge weights of the low-energy states over three edges of this system. This can be written as $w_1(E) = w_2(E) = w_3(E)$ for $|E|<0.25\,\mathrm{eV}$.

The armchair triangular QD in the semi-Dirac limit exhibits different $A_{kl}(E)$ profiles in the extended energy range $-4\,\mathrm{eV}\leq E\leq 4\,\mathrm{eV}$. The distributions explicitly evolve from weak to strong energy-dependent profiles induced by the anisotropic hopping ratio. As shown in Fig.~\ref{fig_5} (b), $A_{12}(E)$ represented by the orange distribution is always pinned at zero across the entire TB energy range. Here, we restrict ourselves to the FL, leading to $w_1(E_F) = w_2(E_F)$. On the other hand, the black and green distributions show sharp vertical accumulations near the FL, as indicated by the colored arrows. The anisotropy values jump to $A_{23}(E_{F}) = +1$ and $A_{31}(E_{F}) = -1$, respectively. Both expressions yield the same result, that is, $w_3(E_{F}) = 0$ (see Eq.~\eqref{PEA}), and these results can be summarized as follows:

\begin{equation}
w_1(E_F)=w_2(E_F), \quad w_3(E_F)=0.
\end{equation}

\begin{figure*}
	\floatbox[{\capbeside\thisfloatsetup{capbesideposition={right,center},capbesidewidth=6.4cm}}]{figure}[\FBwidth]
	{\caption{Spatially-resolved spin densities calculated within the extended MFH model for armchair triangular QDs with $N_{\mathrm{edge}} = 9$. The left and right panels correspond to $t_2/t_1=2$ and $t_2/t_1=2.5$, respectively. The red (blue) circles denote positive (negative) spin polarization localized on the A (B) type sublattice. In both cases, the ground-state total spin is zero ($S_z=0$), corresponding to AFM coupling between two magnetically active edges.}\label{fig_6}}
	{\includegraphics[width=4.25in]{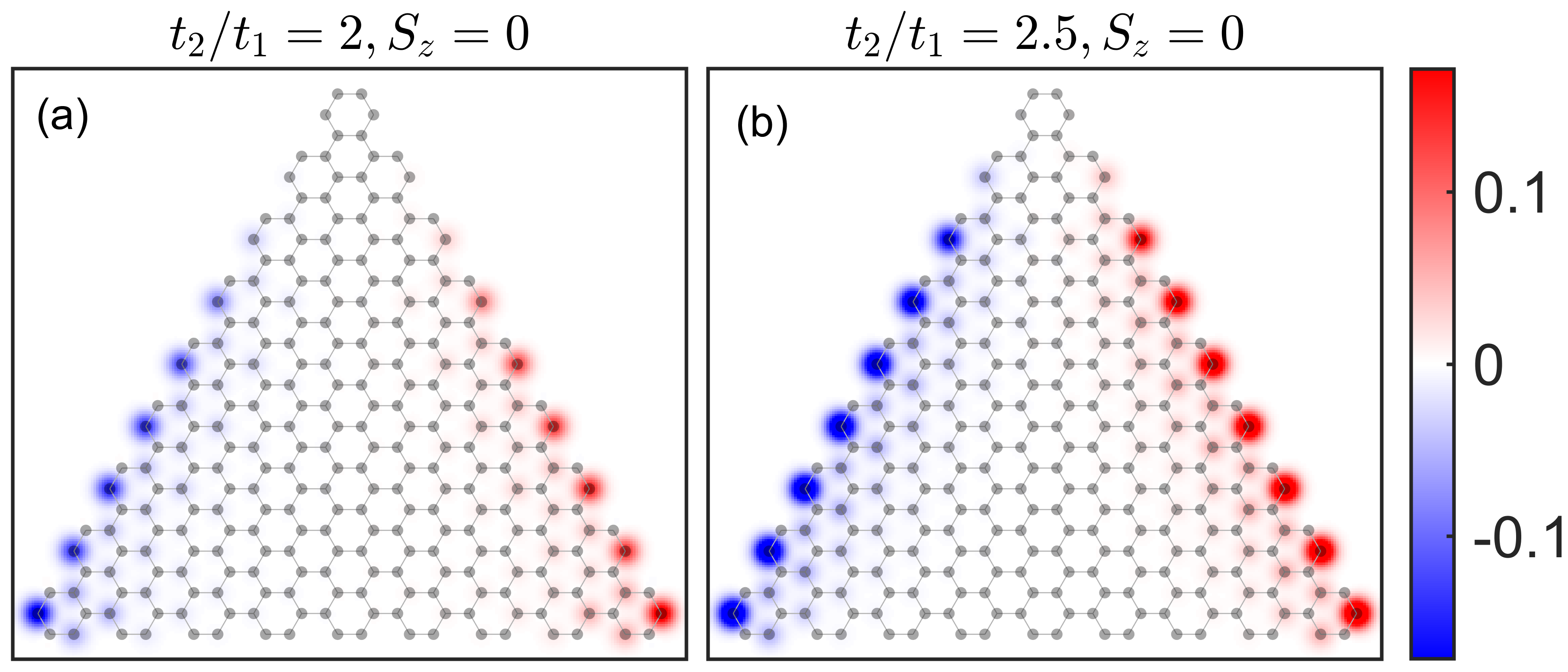}}
\end{figure*}

\begin{figure}[b]
	\begin{center}
		\includegraphics[width=1\linewidth, height=\textheight,keepaspectratio]{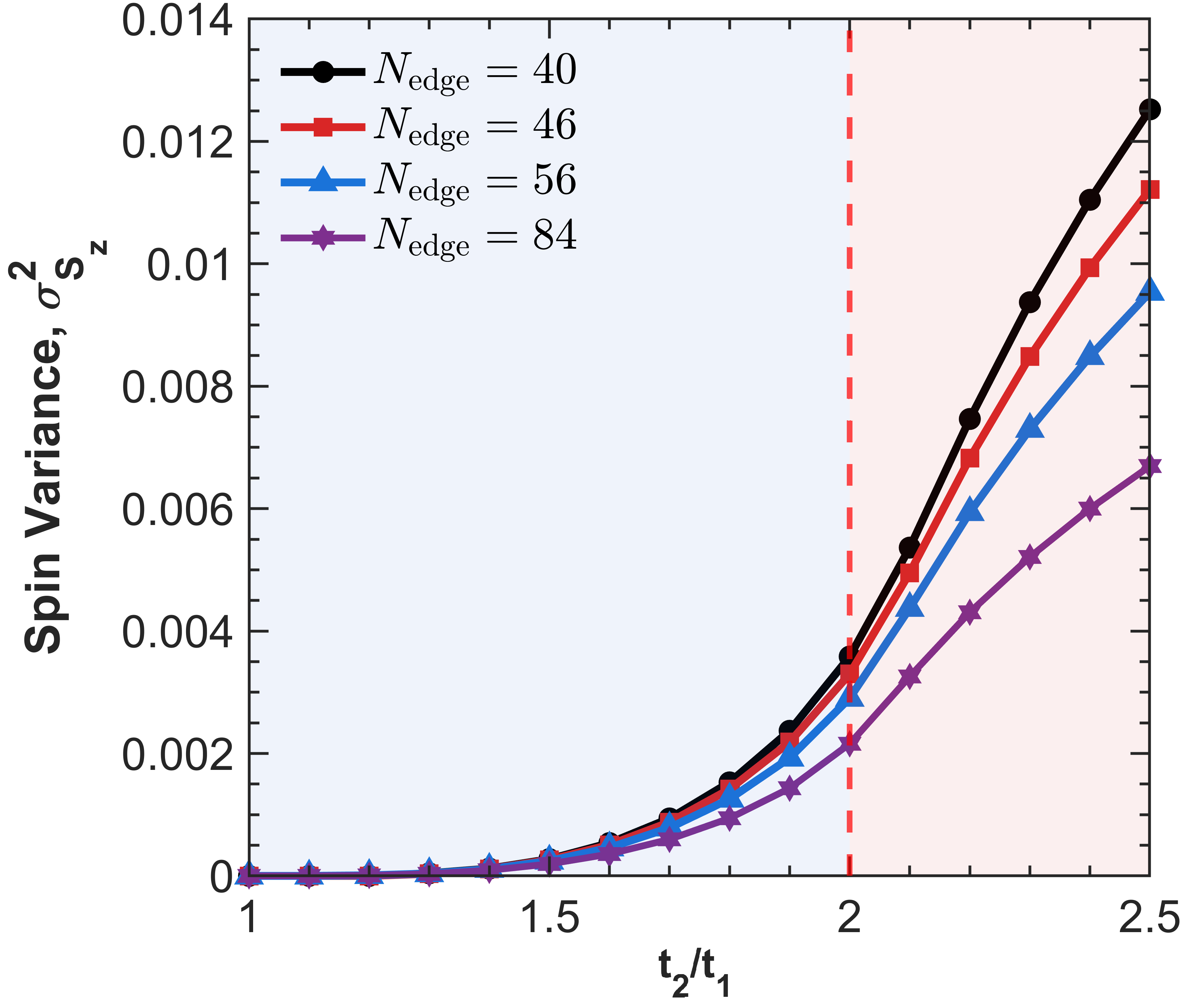}
		\caption{Spin variance, $\sigma_{S_z}^{2}$, as a function of $t_2/t_1$ for the armchair triangular QDs with $N_{\mathrm{edge}}=40$, $46$, $56$, and $84$. The dashed vertical line at $t_2/t_1=2$ separates the uniform and nonuniform regions. The spin variance remains nearly zero in the $t_2/t_1 < 2$ regime and increases rapidly when $t_2/t_1 > 2$. For a fixed hopping anisotropy, the spin variance is inversely proportional to the system size.}
\label{fig_7} 
	\end{center}
\end{figure}

These three edge weights suggest that the states near the FL are equally localized only two edges of the armchair triangular QD in the semi-Dirac limit instead of three edges observed in the Dirac limit. Away from the FL, both $A_{23}(E)$ and $A_{31}(E)$ display inhomogeneous distribution across all energies in the right panel. As the energy varies, they form highly structured clusters, collectively creating a butterfly-like pattern. These clusters remain the same under energy inversion, corresponding to the electron-hole symmetry in the semi-Dirac limit. The anisotropy values approach unity near $|E| = 2\,\mathrm{eV}$, indicating the existence of non-trivial correlations between energy and edge localization at intermediate energies. 

The pairwise edge anistropy results mimic the presence of an unconventional edge magnetism in the armchair triangular QDs for the larger values of $t_2/t_1$. Therefore, we examine ground state properties of these QDs using the extended MFH model. In this context, a smaller system with $N_{\mathrm{edge}} = 9$ is chosen to clearly demonstrate the spatial distribution of spin densities. At $t_2/t_1 = 1$, the ground state is nonmagnetic, which corresponds to the well-known armchair triangular graphene QDs (not shown here) \cite{gucclu2014graphene,polat2020electronic}. The absence of edge magnetism in this case originates from the lack of zero-energy edge states and is consistent with our previous finding obtained from the TB method that is $w_1(E)=w_2(E)=w_3(E)$ for $|E| < 0.25\,\mathrm{eV}$. 

Here, we study spatially-resolved spin densities of armchair triangular QDs with $N_{\mathrm{edge}} = 9$ at $t_2/t_1 = 2$ (left panel) and $t_2/t_1 = 2.5$ (right panel), as shown in Fig.~\ref{fig_6} (a)-(b), respectively. The red (blue) circles correspond to positive (negative) spin polarization localized on the A (B) type sublattice. The localized magnetic moments exhibit FM ordering only along two armchair edges, whereas the bottom edge remains nonmagnetic at $t_2/t_1 = 2$. The ground-state total spin is zero, $S_z=0$, due to the AFM coupling between two magnetically active edges. This distribution shows an edge-asymmetric AFM ground state in the semi-Dirac limit and is in agreement with our previous edge weight results in this limit, which are $w_1(E_F) = w_2(E_F)$ and $w_3(E_{F}) = 0$. Our results are also consistent with the edge-to-edge AFM ground state identified in neutral armchair graphene NRs under shear strain \cite{li2024armchair} and the zero-energy states predicted for armchair edge graphene due to the anisotropic hopping ratio \cite{dahal2010edge}. An anisotropy-induced spatial reconstruction of the spin density becomes more pronounced at $t_2/t_1 = 2.5$, as shown in Fig.~\ref{fig_6} (b).

\begin{figure*}[t]
   \centering
    \includegraphics[width=1\textwidth]{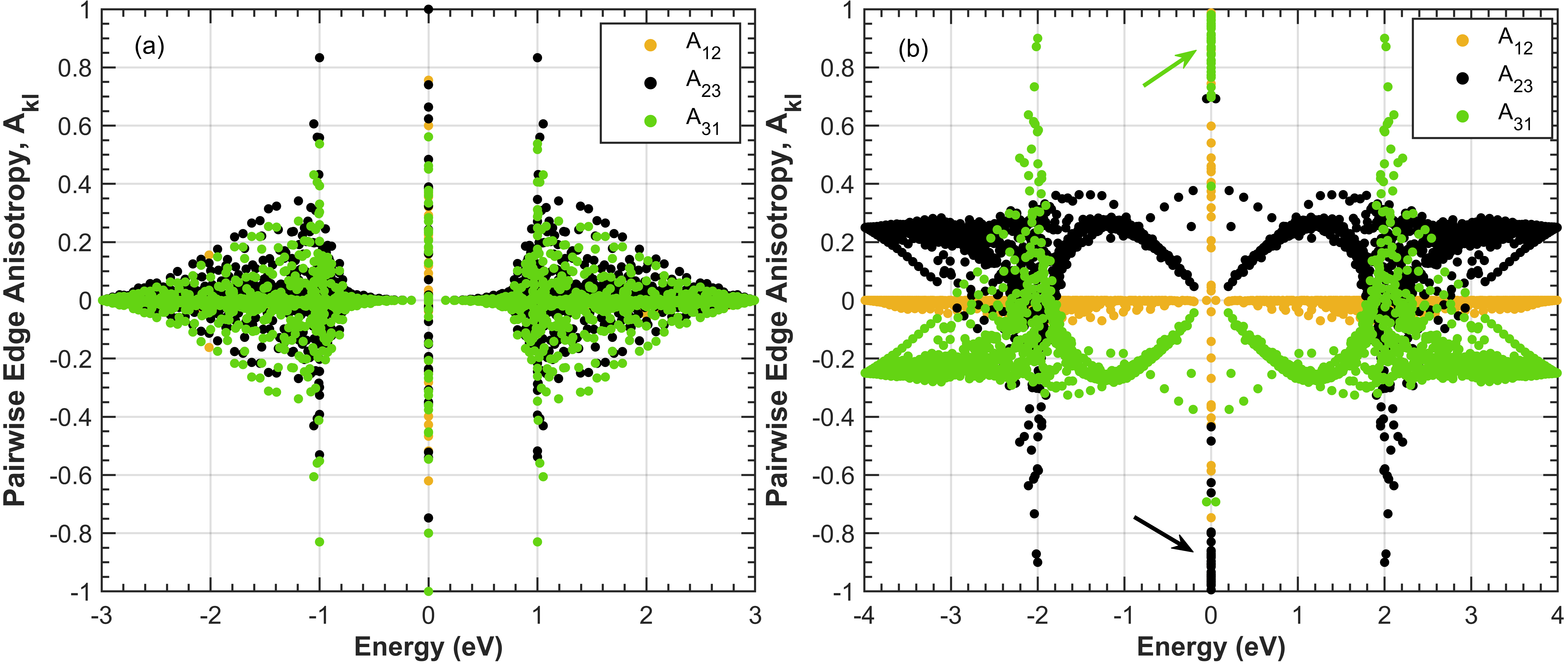}
    \caption{The pairwise edge anisotropies as a function of energy as in Fig.~\ref{fig_5}, but now for the zigzag triangular QDs with $N_{\mathrm{edge}} = 37$. In the right panel, there is abrupt jumps near the FL, marked by the colored arrows, indicating the presence of low-energy states localized only on the bottom edge of this system.}
\label{fig_8}
\end{figure*}

The effects of system size on the ground state properties of armchair triangular QDs are studied with the help of spin variance, $\sigma_{S_z}^{2}$. Fig.~\ref{fig_7} shows the evolution of $\sigma_{S_z}^{2}$ as a function of $t_2/t_1$ for the systems with $N_{\mathrm{edge}} = 40$, $46$, $56$, and $84$. For all considered system sizes, $\sigma_{S_z}^{2}$ remains close to zero and is nearly negligible in the $t_2/t_1<2$ regime. These spin variances indicate that the local spin polarization has an almost uniform distribution across these systems and exhibits negligibly small deviations from its spatial average. 

In contrast, a pronounced increase emerges in the $t_2/t_1 > 2$ regime due to an anisotropy-induced redistribution of the spin densities. In this regime, $\sigma_{S_z}^{2}$ rises monotonically in all system sizes, corresponding to an enhanced spatial inhomogeneity. Furthermore, the magnitude of the spin variance is inversely proportional to the system size and is gradually suppressed for larger systems due to the increasing number of bulk atoms. These results reveal that a spontaneous symmetry breaking occurs in these systems, pointing to an edge-asymmetric spin density pattern. It can be noted that the spatially-resolved spin densities of these investigated QDs were also studied and were found to exhibit the same features as those shown in Fig.~\ref{fig_6} (a)-(b), but not shown here.
\subsection{Zigzag Triangular QDs}
Up to now, we study the armchair triangular QDs, but the zigzag edge termination is not considered. The edge termination plays a critical role in determining the electronic and magnetic properties of the QDs. As previously reported, the sublattice imbalance originating from the zigzag edge termination leads to the formation of FM ground state in the zigzag triangular graphene QDs, consistent with Lieb's rule \cite{kul2025quantum}. In the light of this fact, we investigate the effects of edge termination through the TB and the extended MFH models. 

We analyze $A_{kl}(E)$ profiles of the zigzag triangular QDs with $N_{\mathrm{edge}} = 37$ in the Dirac and the semi-Dirac limits, and the results are shown in Fig.~\ref{fig_8} (a)-(b), respectively. Here, the TB method is utilized to calculate $A_{12}(E)$, $A_{23}(E)$, and $A_{31}(E)$, which are represented by orange, black, and green distributions, respectively. These distributions are mainly centered around zero over the full energy range in the Dirac limit, as shown in Fig.~\ref{fig_8} (a). Their spread is strongly clustered, and a significant number of data points satisfy $|A_{ij}(E)| < 0.2$. The most striking feature of these distributions is the vertical accumulations at the FL, where the pairwise edge anisotropies span almost the entire range of $[-1,1]$ at $t_2/t_1 = 1$. This is due to the well-known zero-energy edge states in the zigzag triangular graphene QDs \cite{gucclu2013zero}. In addition, the black and green distributions with a diamond-like envelope gradually decay and reach zero at the band edges. Here, the orange distribution is not clearly visible, as it lies beneath the black and green distributions. The orange distribution is nearly zero away from the FL, suggesting that the edge asymmetry may also appear in this system, especially at $\lvert E \rvert \approx 1\,\mathrm{eV}$. As can be seen from Fig.~\ref{fig_8} (a)-(b), all solutions are symmetric with respect to $E = 0\,\mathrm{eV}$ due to the electron-hole symmetry in both limits. 

On the other hand, the distributions are well-structured in the semi-Dirac limit and exhibit a strong energy dependence across the entire energy range $-4\,\mathrm{eV} \leq E \leq 4\,\mathrm{eV}$. The most prominent feature is the sharp vertical accumulations of the black and green data points at $E = 0\,\mathrm{eV}$, leading to a dense population near the two maxima, as indicated by the colored arrows in Fig.~\ref{fig_8} (b). These data points mostly reach the values satisfying $\lvert A_{ij} \rvert > 0.8$; therefore, we approximate them as $A_{23}(E_{F}) = -1$ and $A_{31}(E_{F}) = +1$, respectively. This is the sign-inverted version of the armchair triangular QD in the semi-Dirac limit (see also Fig.~\ref{fig_5} (b)). A straightforward algebra under this approximation can prove the following statement

\begin{equation}
w_1(E_F) = w_2(E_F) \approx 0, w_3(E_F) \neq 0.
\end{equation}

This means that the three edges of the triangle do not behave similarly, and the electronic structure intrinsically develops an asymmetric localization pattern, where the weights appear only along the bottom edge of the triangular geometry. In this manner, these results can be an early indication of the emergence of edge-asymmetric FM ground state in the zigzag triangular QDs in the semi-Dirac limit. 

These black and green distributions exhibit well-defined curved branches away from the FL. The black distribution forms arch-like structures centered around $\lvert E \rvert \approx 1\,\mathrm{eV}$, and the anisotropy values mostly spread in the range of $0.2 < A_{23}(E) < 0.4$. The green distribution shows a complementary behavior. The negative anisotropy values mostly lie within the range of $-0.2 < A_{31}(E) < -0.4$, and there are two symmetric lobes around $E = 0\,\mathrm{eV}$. The zigzag triangular QD in the semi-Dirac limit has more regular branches, in contrast to the more scattered clusters emerged in the armchair triangular QD in the same limit. As for the orange distribution, it remains nearly fixed around $A_{12}(E) \approx 0$ away from the FL, indicating the presence of a correlation between energy and edge localization, especially around $|E| = 2\,\mathrm{eV}$.

\begin{figure}[t]
	\begin{center}
		\includegraphics[width=1\linewidth, height=\textheight,keepaspectratio]{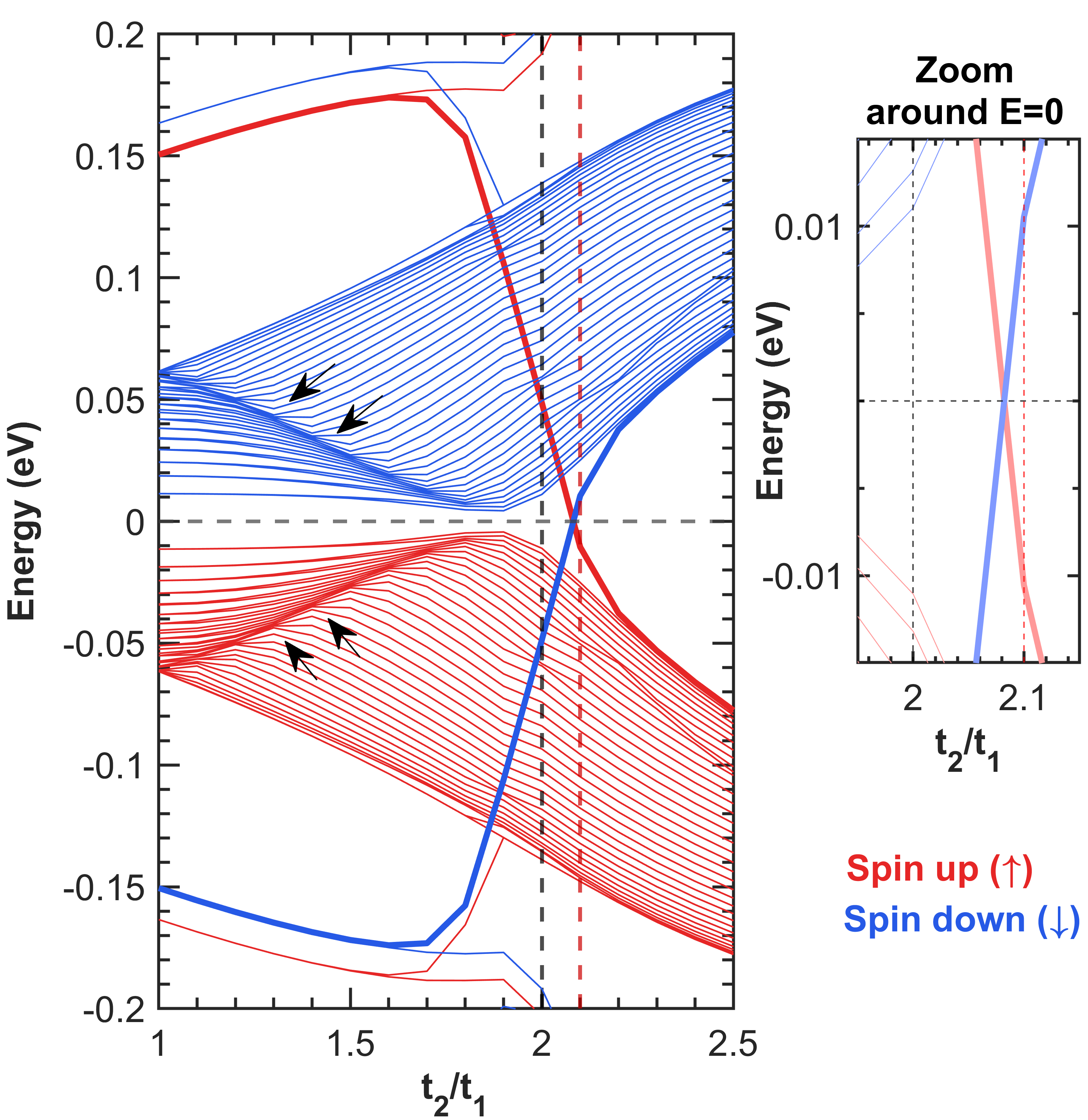}
		\caption{Spin-dependent spectral evolution of low-energy states as a function of $t_2/t_1$ for the zigzag triangular QDs with $N_{\mathrm{edge}} = 37$. In the $t_2/t_1 < 2$ regime, the black arrows show some of avoided crossings between the edge states. In the $t_2/t_1 > 2$ regime, there is a genuine level crossing, which is clearly shown in the zoomed-in view in the right panel. Here the black and red vertical dashed lines marks $t_2/t_1 = 2$ and $t_2/t_1 = 2.1$, respectively.}
\label{fig_9} 
	\end{center}
\end{figure}

In the presence of electron-electron interactions, the zigzag triangular QDs with $N_{\mathrm{edge}} = 37$ develops a FM ground state. The low-energy electronic states as a function of $t_2/t_1$ are shown in Fig.~\ref{fig_9}. In the $t_2/t_1 < 2$ regime, there are numerous avoided crossings within each spin sector, and some of them are marked by the black arrows in this spectrum. These avoided crossings indicate that the edge states interact with each other through the level repulsion according to the von Neumann‐Wigner non-crossing principle \cite{landau1958quantum}. This interaction results in hybridized edge states and modifies their spatial distribution characteristics. As a result, the edge-asymmetric magnetization starts to appear, but is weak in this regime, as discussed in the following section. 

As is clear from Fig.~\ref{fig_9}, there are twofold-degenerate bulk-like states in the spectrum, located above and below the FL. Starting from $t_2/t_1 > 1.6$, the degeneracies of these energy levels are lifted on both electron and hole channels. One electronic state from each pair simultaneously approaches to the FL (the two thickened lines), whereas the others gradually move away from the FL. As is clear from the zoomed-in view in the right panel, these two approaching states from different spin sectors cross each other at $t_2/t_1 \approx 2$. As a result, a bulk-like hole state in the spin-up spectrum turns into an edge-like electron state, whereas a bulk-like electron state in the spin-down spectrum becomes an edge-like hole state. This genuine level crossing indicates a transition induced by the increasing hopping anisotropy. At this point, it is useful to remember the Lieb sublattice imbalance rule, which predicts the ground-state total spin as $S_z$ = ($N_{\mathrm{edge}}$ - 1)/2 in the interacting case \cite{fernandez2007magnetism}. Owing to this transition, the ground-state total spin of the analyzed system is expected to shift from $S_z$ to $S_z + 1$ for $t_2/t_1 > 2$. This is due to the interplay between hopping anisotropy, electron-electron interactions, and the complete quantum confinement, rather than a failure of Lieb’s rule. As a result, the ground-state total spin of this system can be written as follows:

\begin{equation}
S_z =
\begin{cases}
(N_{\mathrm{edge}}-1)/2, & t_2/t_1 \leq 2,\\
(N_{\mathrm{edge}}+1)/2, & t_2/t_1 > 2.
\end{cases}
\end{equation}

\begin{figure}[b]
\centering
\includegraphics[width=\linewidth,keepaspectratio]{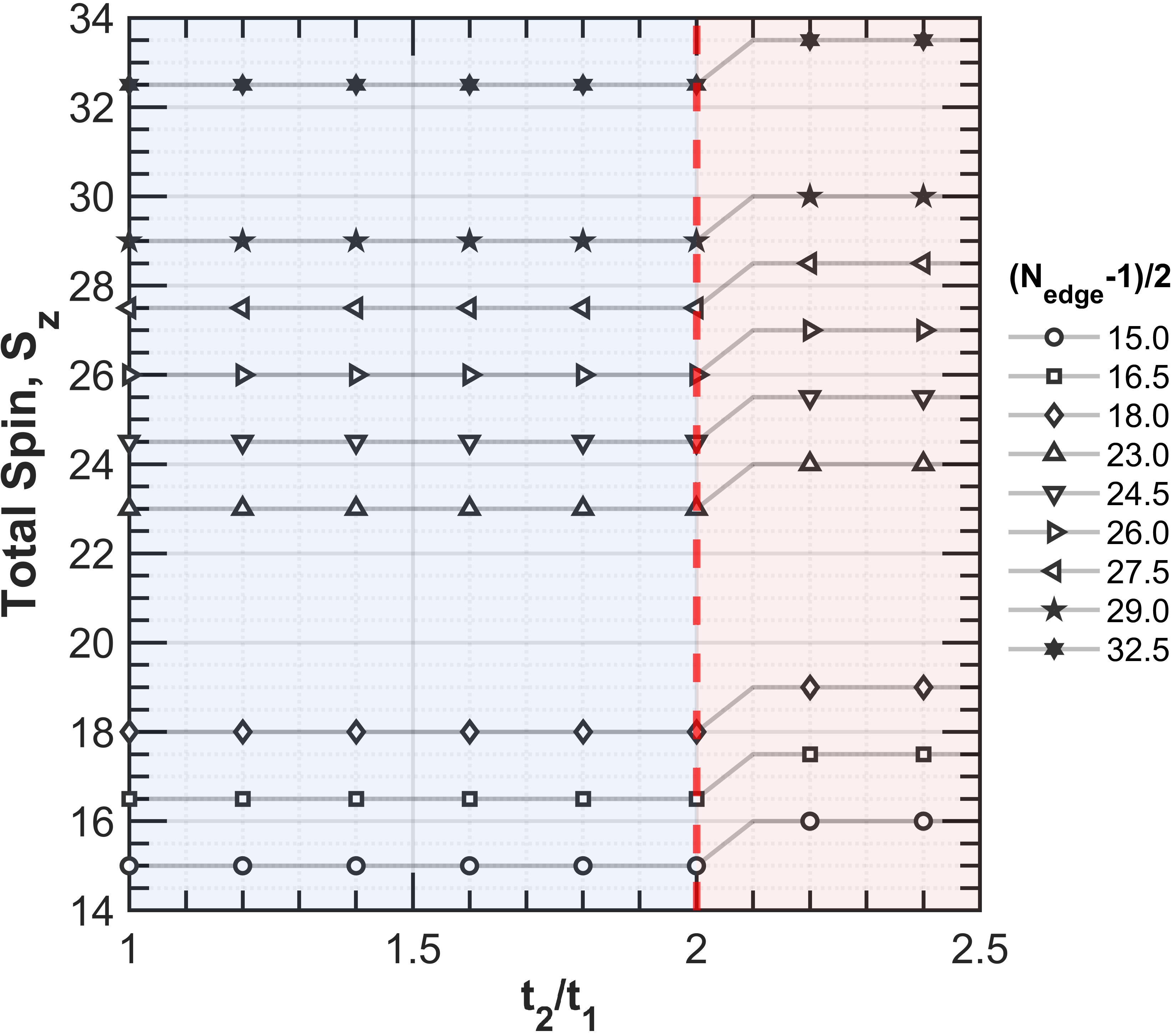}
\caption{
The ground-state total spin $S_z$ as a function of $t_2/t_1$ for the systems with $N_{\mathrm{edge}} = 31$ to $66$. A spin plateau emerges in the $t_2/t_1 < 2$ regime, corresponding to Lieb’s rule. For all sizes, it reaches $S_z + 1$ in the $t_2/t_1 > 2$ regime due to the genuine level crossing in each case.}
\label{fig_10}
\end{figure}

\begin{figure*}[t]
   \centering
    \includegraphics[width=1\textwidth]{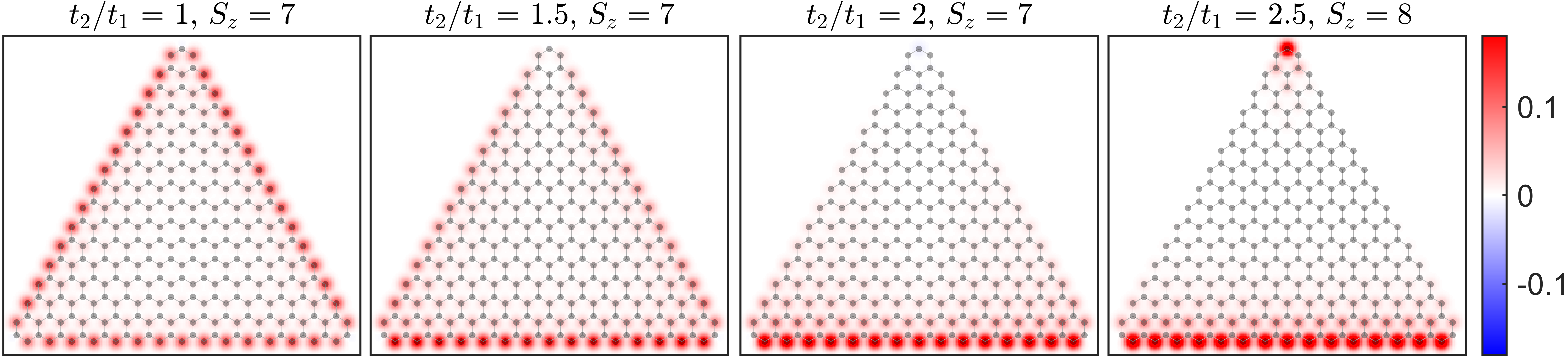}
    \caption{The spatially-resolved spin densities of the zigzag triangular QDs with $N_{\mathrm{edge}} = 15$ at the selected values of $t_2/t_1 = 1, 1.5,2$, and $2.5$, from left to right. A total increase of $\Delta S_z = +1$ occurs at $t_2/t_1 = 2.5$, where the spin-up density exhibits an unusual localization on the B-type atom located at the top of triangular geometry.}
\label{fig_11}
\end{figure*}

The zigzag triangular QDs with various system sizes ranging from $N_{\mathrm{edge}} = 9$ to $100$ are studied, and the effects of finite-size on the magnetic ground state properties are discussed in greater detail. A previous study has shown that adding a single electron leads to the complete spin depolarization in the zigzag triangular graphene QDs with $N_{\mathrm{edge}} \leq 9$ \cite{potasz2012electronic}. In this context, all system sizes considered here are mostly larger than this critical size, and our results are independent of this small-size effect. 

The ground-state total spin $S_z$ as a function of $t_2/t_1$ are shown in Fig.~\ref{fig_10} for various system sizes. For the sake of simplicity, we only show the ground-state total spin of the systems with $N_{\mathrm{edge}} = 31$ to $66$. As is clear from Fig.~\ref{fig_10}, there are two distinct regimes with respect to $t_2/t_1$. These systems exhibit a constant $S_z$ in the $t_2/t_1 < 2$ regime and develop an extended spin plateau despite the progressive increasing of $t_2/t_1$. In this regime, the ground-state total spin is equivalent to that of the Dirac limit and is protected against weak hopping anisotropy.

These systems exhibit qualitatively different behavior for $t_2/t_1 > 2$ in our extended MFH calculations. All triangular geometries enter the reconstructed spin regime, and the ground-state total spin spontaneously increases by exactly one unit, resulting in a total increase of $\Delta S_z = +1$ due to the genuine level crossing (see also the zoomed-in view in Fig.~\ref{fig_9}). In this context, a step-like evolution of the total spin is prominent in the ground states of these considered systems, as is clear from Fig.~\ref{fig_10}. The exact same behavior is also observed for the sample sizes up to $N_{\mathrm{edge}} = 100$, which is the largest system under investigation. 

We observe that the larger system sizes sharpens the transition near the value $t_2/t_1 = 2$, but the smaller QDs with $N_{\mathrm{edge}} < 31$ delay this instability by requiring slightly larger values of $t_2/t_1$. However, the transition discussed here also occurs across all considered system sizes. To demonstrate the edge-asymmetric FM ground state, we study the spatially-resolved spin densities of a smaller system size with $N_{\mathrm{edge}} = 15$ at the selected values of $t_2/t_1$, as shown in Fig.~\ref{fig_11} (a)-(d). The spatially-resolved spin density is uniform though all edges of the analyzed system at $t_2/t_1 = 1$, corresponding to the zigzag triangular graphene QDs \cite{han2024zigzag,hagymasi2018entanglement}. When the anisotropic hopping ratio reaches to $t_2/t_1 = 1.5$, an enhanced weight of spin density on the bottom edge of the system emerges due the hybridized edge states. In this context, the edge-asymmetric magnetization is more pronounced at $t_2/t_1 = 2$. It is clear that the remaining two zigzag boundaries and the interior region remain nearly spin-compensated. In addition, the ground-state total spin of the analyzed system remains constant at $S_z = 7$. 

The spatially-resolved spin density demonstrates that the system stabilizes the single edge magnetism at $t_2/t_1 = 2.5$. The magnetic ground state evolves from a geometrically equivalent edge configuration toward a single edge configuration ground state with increasing $t_2/t_1$. The ground-state total spin equals to $S_z =8$ for $N_{\mathrm{edge}} = 15$ at $t_2/t_1 = 2.5$. On the other hand, the spin-up density exhibits an unusual localization on the B-type atom located at the top of triangular geometry, leading to a reversal of the usual sublattice polarization.

In the context of edge-asymmetric magnetism, recent experimental and theoretical research on Janus graphene NRs has confirmed the existence of a FM ground state localized along the pristine zigzag edge, whereas magnetism is completely quenched at the opposite defective edge \cite{song2025janus}. This finding reveals an effective approach for the edge asymmetry in graphene-based systems. In this context, our results indicate dispersion-driven mechanism for realizing single-edge ferromagnetism in finite triangular structures, demonstrating that anisotropic hopping and the associated transition can play a crucial role in determining the spatial distribution of magnetic ordering.
\subsection{Armchair and Zigzag Hexagonal QDs}
\begin{figure}[t]
	\begin{center}
		\includegraphics[width=1\linewidth, height=\textheight,keepaspectratio]{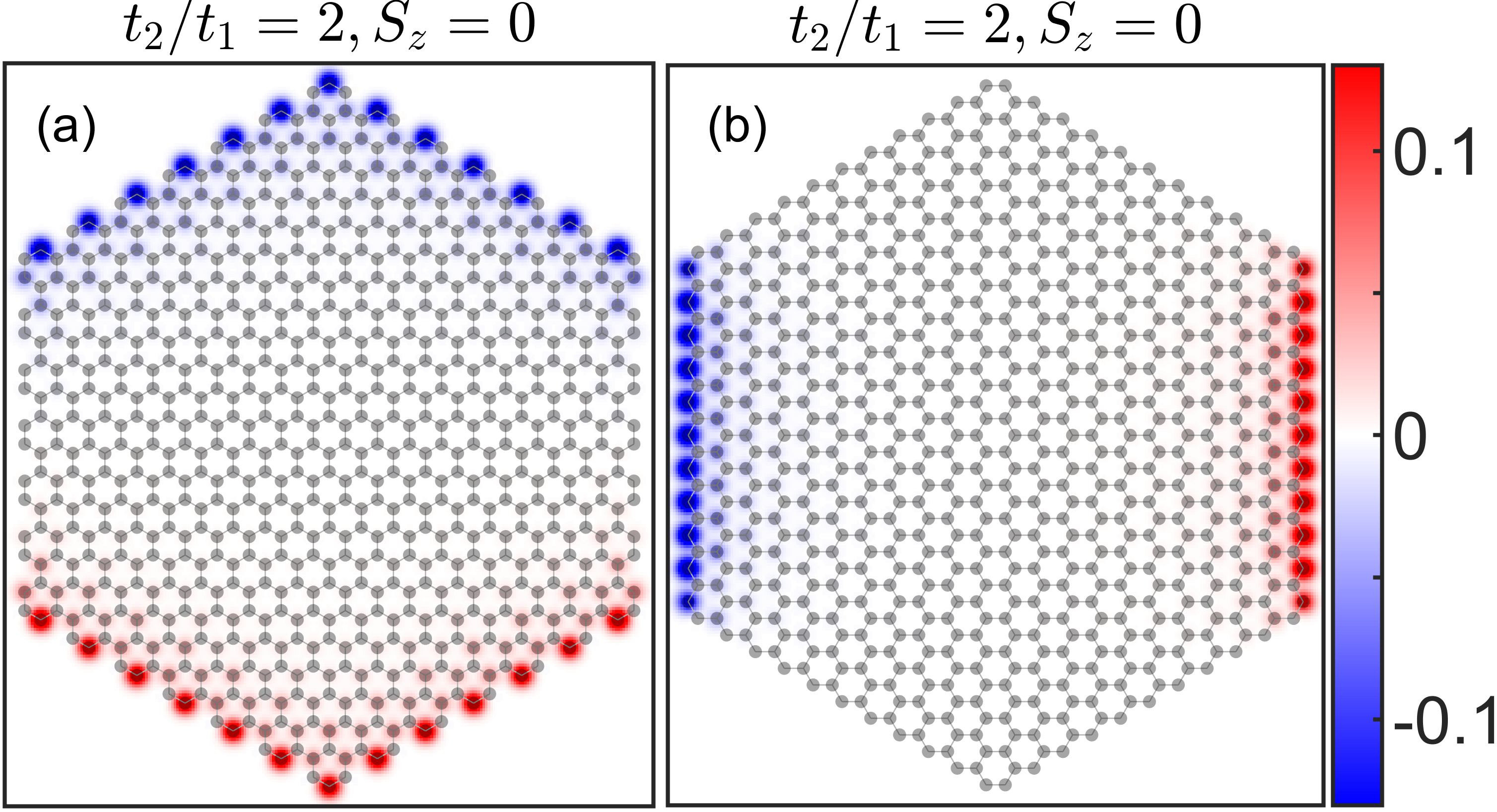}
		\caption{Spatially-resolved spin densities of (a) armchair hexagonal QD with $N_{\mathrm{edge}} = 7$ and (b) zigzag hexagonal QD with $N_{\mathrm{edge}} = 11$ at $t_2/t_1=2$. In both systems, the ground state exhibits edge-asymmetric AFM ordering in the semi-Dirac limit.}
\label{fig_12}
	\end{center}
\end{figure}

We further study the spatially-resolved spin densities of hexagonal QDs \cite{polat2020collapse,polat2020atomic,kul2020electronic}. As shown in Fig.~\ref{fig_12} (a), the armchair hexagonal QD with $N_{\mathrm{edge}} = 7$ develops an edge-asymmetric AFM ground state at $t_2/t_1 = 2$. In particular, the spins are ferromagnetically aligned throughout the upper and lower edges of the system, and the ground-state total spin is zero. Here, the A and B sublattices host the spin-up and spin-down components, respectively. 

As is clear from Fig.~\ref{fig_12} (b), the ground state exhibits an edge-asymmetric AFM ordering in the zigzag hexagonal QD with $N_{\mathrm{edge}} = 11$. The system have a FM ordering along only two opposite zigzag edges, instead of spreading across all six edges \cite{hu2019room}. This is due to the underlying semi-Dirac electronic structure, and the anisotropic hopping amplitudes decide the magnetically active edges. These results indicate that the edge-asymmetric magnetic configurations are not unique to the triangular QDs, and the anisotropic hopping ratio also alters the electronic and magnetic properties of the hexagonal QDs. In our test calculations performed with the smaller values of $U$, as well as with $V_{\mathrm{off}} = 0$, we found a similar edge-asymmetric AFM ground state in the zigzag hexagonal QDs with $N_{\mathrm{edge}} = 11$, revealing the same qualitative physical behavior and the robustness of the edge-asymmetric ground state.
\section{\label{sec:conclusions} CONCLUSION}
The 2D semi-Dirac systems possess an anisotropic dispersion relation, leading to physical phenomena that have no direct counterpart in either massless Dirac fermions or conventional massive charge carriers. In this study, we investigate 0D signatures of 2D semi-Dirac physics through the QDs with various sizes, shapes, and edge terminations. 

In the case of non-interacting semi-Dirac fermions, the energy-resolved pairwise anisotropies indicate that the low-energy electronic states exhibit asymmetric patterns of localization among the three edges of triangular QDs. Proceeding from this point, we systematically perform the self-consistent calculations throughout the transition from the Dirac limit to the semi-Dirac limit and beyond. Hence, we identify edge-asymmetric AFM and FM ground states in the armchair and zigzag triangular QDs, respectively. In particular, the magnetic moments are localized only along the bottom edge of the zigzag triangular QDs, which becomes more pronounced in the semi-Dirac limit and beyond. We also show the presence of a high-spin phase in the $t_2/t_1 > 2$ regime for the zigzag triangular QDs of various sizes. These results indicate that the anisotropic hopping ratio acts as a tuning parameter and controls the ground-state total spin of zigzag triangular QDs. On the other hand, the remaining two edges are antiferromagnetically coupled in the case of armchair triangular QDs. 

The results of edge-asymmetric AFM ground state in the hexagonal QDs further indicate that the edge-asymmetric magnetic configurations are not unique to the triangular QDs. Overall, the edge-asymmetric magnetism could create potential opportunities for future spintronics and the next-generation devices. 
\vspace*{1\baselineskip}
\begin{acknowledgments}
The author received no external funding for this work and is grateful to Prof. Dr. Alev Devrim Güçlü (Department of Physics, İzmir Institute of Technology) and Prof. Dr. Haldun Sevinçli (Department of Physics, Bilkent University) for their valuable discussions. The author also sincerely thanks his colleagues at Istanbul Health and Social Sciences Vocational School for providing a supportive academic environment.
\end{acknowledgments}
\bibliographystyle{apsrev4-2} (no title)
\bibliography{Ref}

@article{novoselov2004electric,
  title={Electric field effect in atomically thin carbon films},
  author={Novoselov, Kostya S and Geim, Andre K and Morozov, Sergei V and Jiang, De-eng and Zhang, Yanshui and Dubonos, Sergey V and Grigorieva, Irina V and Firsov, Alexandr A},
  journal={Science},
  volume={306},
  number={5696},
  pages={666--669},
  year={2004},
  publisher={American Association for the Advancement of Science}
}

@article{geim2007rise,
  title={The rise of graphene},
  author={Geim, Andre K and Novoselov, Konstantin S},
  journal={Nat. Mater.},
  volume={6},
  number={3},
  pages={183--191},
  year={2007},
  publisher={Nature Publishing Group UK London}
}

@article{castro2009electronic,
  title={The electronic properties of graphene},
  author={Castro Neto, Antonio H and Guinea, Francisco and Peres, Nuno MR and Novoselov, Kostya S and Geim, Andre K},
  journal={Rev. Mod. Phys.},
  volume={81},
  number={1},
  pages={109--162},
  year={2009},
  publisher={APS}
}

@article{watanabe2004direct,
  title={Direct-bandgap properties and evidence for ultraviolet lasing of hexagonal boron nitride single crystal},
  author={Watanabe, Kenji and Taniguchi, Takashi and Kanda, Hisao},
  journal={Nat. Mater.},
  volume={3},
  number={6},
  pages={404--409},
  year={2004},
  publisher={Nature Publishing Group UK London}
}

@article{dean2010boron,
  title={Boron nitride substrates for high-quality graphene electronics},
  author={Dean, Cory R and Young, Andrea F and Meric, Inanc and Lee, Chris and Wang, Lei and Sorgenfrei, Sebastian and Watanabe, Kenji and Taniguchi, Takashi and Kim, Phillip and Shepard, Kenneth L and others},
  journal={Nat. Nanotechnol.},
  volume={5},
  number={10},
  pages={722--726},
  year={2010},
  publisher={Nature Publishing Group}
}

@article{cassabois2016hexagonal,
  title={Hexagonal boron nitride is an indirect bandgap semiconductor},
  author={Cassabois, Guillaume and Valvin, Pierre and Gil, Bernard},
  journal={Nat. Photonics},
  volume={10},
  number={4},
  pages={262--266},
  year={2016},
  publisher={Nature Publishing Group},
  doi={10.1038/nphoton.2015.277}
}

@article{mak2010atomically,
  title={Atomically thin MoS 2: a new direct-gap semiconductor},
  author={Mak, Kin Fai and Lee, Changgu and Hone, James and Shan, Jie and Heinz, Tony F},
  journal={Phys. Rev. Lett.},
  volume={105},
  number={13},
  pages={136805},
  year={2010},
  publisher={APS}
}

@article{splendiani2010emerging,
  title={Emerging photoluminescence in monolayer MoS2},
  author={Splendiani, Andrea and Sun, Liang and Zhang, Yuanbo and Li, Tianshu and Kim, Jonghwan and Chim, Chi-Yung and Galli, Giulia and Wang, Feng},
  journal={Nano Lett.},
  volume={10},
  number={4},
  pages={1271--1275},
  year={2010},
  publisher={ACS Publications}
}

@article{wang2012electronics,
  title={Electronics and optoelectronics of two-dimensional transition metal dichalcogenides},
  author={Wang, Qing Hua and Kalantar-Zadeh, Kourosh and Kis, Andras and Coleman, Jonathan N and Strano, Michael S},
  journal={Nat. Nanotechnol.},
  volume={7},
  number={11},
  pages={699--712},
  year={2012},
  publisher={Nature Publishing Group UK London}
}

@article{chhowalla2013chemistry,
  title={The chemistry of two-dimensional layered transition metal dichalcogenide nanosheets},
  author={Chhowalla, Manish and Shin, Hyeon Suk and Eda, Goki and Li, Lain-Jong and Loh, Kian Ping and Zhang, Hua},
  journal={Nat. Chem.},
  volume={5},
  number={4},
  pages={263--275},
  year={2013},
  publisher={Nature Publishing Group UK London}
}

@article{manzeli20172d,
  title={2D transition metal dichalcogenides},
  author={Manzeli, Sajedeh and Ovchinnikov, Dmitry and Pasquier, Diego and Yazyev, Oleg V and Kis, Andras},
  journal={Nat. Rev. Mater.},
  volume={2},
  number={8},
  pages={17033},
  year={2017},
  publisher={Nature Publishing Group}
}

@article{cahangirov2009two,
  title={Two-and one-dimensional honeycomb structures of silicon and germanium},
  author={Cahangirov, Seymur and Topsakal, Mehmet and Akt{\"u}rk, Ethem and {\c{S}}ahin, Hasan and Ciraci, Salim},
  journal={Phys. Rev. Lett.},
  volume={102},
  number={23},
  pages={236804},
  year={2009},
  publisher={APS}
}

@article{vogt2012silicene,
  title={Silicene: compelling experimental evidence for graphenelike two-dimensional silicon},
  author={Vogt, Patrick and De Padova, Paola and Quaresima, Claudio and Avila, Jose and Frantzeskakis, Emmanouil and Asensio, Maria Carmen and Resta, Andrea and Ealet, B{\'e}n{\'e}dicte and Le Lay, Guy},
  journal={Phys. Rev. Lett.},
  volume={108},
  number={15},
  pages={155501},
  year={2012},
  publisher={APS}
}

@article{xu2013large,
  title={Large-gap quantum spin Hall insulators in tin films},
  author={Xu, Yong and Yan, Binghai and Zhang, Hai-Jun and Wang, Jing and Xu, Gang and Tang, Peizhe and Duan, Wenhui and Zhang, Shou-Cheng},
  journal={Phys. Rev. Lett.},
  volume={111},
  number={13},
  pages={136804},
  year={2013},
  publisher={APS}
}

@article{molle2017buckled,
  title={Buckled two-dimensional Xene sheets},
  author={Molle, Alessandro and Goldberger, Joshua and Houssa, Michel and Xu, Yong and Zhang, Shou-Cheng and Akinwande, Deji},
  journal={Nat. Mater.},
  volume={16},
  number={2},
  pages={163--169},
  year={2017},
  publisher={Nature Publishing Group UK London}
}

@article{liu2014phosphorene,
  title={Phosphorene: an unexplored 2D semiconductor with a high hole mobility},
  author={Liu, Han and Neal, Adam T and Zhu, Zhen and Luo, Zhe and Xu, Xianfan and Tom{\'a}nek, David and Ye, Peide D},
  journal={ACS nano},
  volume={8},
  number={4},
  pages={4033--4041},
  year={2014},
  publisher={ACS Publications}
}

@article{li2014black,
  title={Black phosphorus field-effect transistors},
  author={Li, Likai and Yu, Yong and Ye, Gui-Jun and Ge, Qingqin and Ou, Xing and Wu, Hui and Feng, Donglai and Chen, Xian-Hui and Zhang, Yuanbo},
  journal={Nat. Nanotechnol.},
  volume={9},
  number={5},
  pages={372--377},
  year={2014},
  publisher={Nature Publishing Group},
  doi={10.1038/nnano.2014.35}
}

@article{carvalho2016phosphorene,
  title={Phosphorene: from theory to applications},
  author={Carvalho, Alexandra and Wang, Min and Zhu, Xi and Rodin, Aleksandr S and Su, Haibin and Castro Neto, Antonio H},
  journal={Nat. Rev. Mater.},
  volume={1},
  number={11},
  pages={16061},
  year={2016},
  publisher={Nature Publishing Group}
}

@article{butler2013progress,
  title={Progress, challenges, and opportunities in two-dimensional materials beyond graphene},
  author={Butler, Sheneve Z and Hollen, Shawna M and Cao, Linyou and Cui, Yi and Gupta, Jay A and Guti{\'e}rrez, Humberto R and Heinz, Tony F and Hong, Seung Sae and Huang, Jiaxing and Ismach, Ariel F and others},
  journal={ACS nano},
  volume={7},
  number={4},
  pages={2898--2926},
  year={2013},
  publisher={ACS Publications}
}

@article{miro2014atlas,
  title={An atlas of two-dimensional materials},
  author={Mir{\'o}, Pere and Audiffred, Martha and Heine, Thomas},
  journal={Chem. Soc. Rev.},
  volume={43},
  number={18},
  pages={6537--6554},
  year={2014},
  publisher={Royal Society of Chemistry}
}

@article{bhimanapati2015recent,
  title={Recent advances in two-dimensional materials beyond graphene},
  author={Bhimanapati, Ganesh R and Lin, Zhong and Meunier, Vincent and Jung, Yeonwoong and Cha, Judy and Das, Saptarshi and Xiao, Di and Son, Youngwoo and Strano, Michael S and Cooper, Valentino R and others},
  journal={ACS nano},
  volume={9},
  number={12},
  pages={11509--11539},
  year={2015},
  publisher={ACS Publications}
}

@article{khan2020recent,
  title={Recent developments in emerging two-dimensional materials and their applications},
  author={Khan, Karim and Tareen, Ayesha Khan and Aslam, Muhammad and Wang, Renheng and Zhang, Yupeng and Mahmood, Asif and Ouyang, Zhengbiao and Zhang, Han and Guo, Zhongyi},
  journal={J. Mater. Chem. C},
  volume={8},
  number={2},
  pages={387--440},
  year={2020},
  publisher={Royal Society of Chemistry}
}

@article{Schaibley2016Valleytronics,
  author  = {Schaibley, John R. and Yu, Hongyi and Clark, Genevieve and Rivera, Pasqual and Ross, Jason S. and Seyler, Kyle L. and Yao, Wang and Xu, Xiaodong},
  title   = {Valleytronics in 2D Materials},
  journal = {Nat. Rev. Mater.},
  volume  = {1},
  pages   = {16055},
  year    = {2016},
  doi     = {10.1038/natrevmats.2016.55}
}

@article{burch2018magnetism,
  title={Magnetism in two-dimensional van der Waals materials},
  author={Burch, Kenneth S and Mandrus, David and Park, Je-Geun},
  journal={Nature},
  volume={563},
  number={7729},
  pages={47--52},
  year={2018},
  publisher={Nature Publishing Group UK London}
}

@article{choudhuri2019recent,
  title={Recent advances in graphene-like 2D materials for spintronics applications},
  author={Choudhuri, Indrani and Bhauriyal, Preeti and Pathak, Biswarup},
  journal={Chem. Mater.},
  volume={31},
  number={20},
  pages={8260--8285},
  year={2019},
  publisher={ACS Publications}
}

@article{liu_naturenano_2020,
  author  = {Liu, Chunsen and Chen, Huawei and Wang, Shuiyuan and Liu, Qi and Jiang, Yu-Gang and Zhang, David Wei and Liu, Ming and Zhou, Peng},
  title   = {Two-Dimensional Materials for Next-Generation Computing Technologies},
  journal = {Nat. Nanotechnol.},
  volume  = {15},
  number  = {7},
  pages   = {545--557},
  year    = {2020},
  doi     = {10.1038/s41565-020-0724-3}
}

@article{vincent2021opportunities,
  title={Opportunities in electrically tunable 2D materials beyond graphene: Recent progress and future outlook},
  author={Vincent, Tom and Liang, Jiayun and Singh, Simrjit and Castanon, Eli G and Zhang, Xiaotian and McCreary, Amber and Jariwala, Deep and Kazakova, Olga and Al Balushi, Zakaria Y},
  journal={Appl. Phys. Rev.},
  volume={8},
  number={4},
  year={2021},
  publisher={AIP Publishing}
}

@article{ozbal2023peculiar,
  author  = {{\"O}zbal Sarg{\i}n, G{\"o}zde and Sarikurt, Sevil and Sevin{\c{c}}li, Haldun and Sevik, Cem},
  title   = {The Peculiar Potential of Transition Metal Dichalcogenides for Thermoelectric Applications: A Perspective on Future Computational Research},
  journal = {J. Appl. Phys.},
  volume  = {133},
  number  = {15},
  pages   = {150901},
  year    = {2023},
  doi     = {10.1063/5.0143458}
}

@article{Gao2018Topological,
title={Topological Semimetals from First Principles},
author={Heng Gao and J. Venderbos and Youngkuk Kim and A. Rappe},
journal={Annu. Rev. Mater. Res.},
year={2018},
doi={10.1146/annurev-matsci-070218-010049}
}

@article{Tokura2019Magnetic,
title={Magnetic topological insulators},
author={Y. Tokura and K. Yasuda and A. Tsukazaki},
journal={Nat. Rev. Phys.},
year={2019},
volume={1},
pages={126 - 143},
doi={10.1038/s42254-018-0011-5}
}

@article{Cayssol2020Topological,
title={Topological and geometrical aspects of band theory},
author={J. Cayssol and J. Fuchs},
journal={J. Phys. Mater.},
year={2020},
volume={4},
doi={10.1088/2515-7639/abf0b5}
}

@article{Bernevig2022Progress,
title={Progress and prospects in magnetic topological materials},
author={B. Bernevig and C. Felser and H. Beidenkopf},
journal={Nature},
year={2022},
volume={603},
pages={41 - 51},
doi={10.1038/s41586-021-04105-x}
}

@article{meng2022hole,
  title={Hole-doping induced ferromagnetism in 2D materials},
  author={Meng, Ruishen and da Costa Pereira, Lino and Locquet, Jean-Pierre and Afanas'ev, Valeri V. and Pourtois, Geoffrey and Houssa, Michel},
  journal={npj Comput. Mater.},
  volume={8},
  pages={230},
  year={2022},
  publisher={Nature Publishing Group},
  doi={10.1038/s41524-022-00916-2}
}

@article{banerjee2009tight,
  title={Tight-binding modeling and low-energy behavior of the semi-Dirac point},
  author={Banerjee, S. and Singh, R. R. P. and Pardo, V. and Pickett, W. E.},
  journal={Phys. Rev. Lett.},
  volume={103},
  number={1},
  pages={016402},
  year={2009},
  publisher={American Physical Society},
  doi={10.1103/PhysRevLett.103.016402}
}

@article{delplace2010semi,
  title={Semi-Dirac point in the Hofstadter spectrum},
  author={Delplace, P and Montambaux, G},
  journal={Phys. Rev. B - Condens. Matter Mater. Phys.},
  volume={82},
  number={3},
  pages={035438},
  year={2010},
  publisher={APS}
}

@article{adroguer2016diffusion,
  title={Diffusion of Dirac fermions across a topological merging transition in two dimensions},
  author={Adroguer, Pierre and Carpentier, David and Montambaux, Gilles and Orignac, Edmond},
  journal={Phys. Rev. B},
  volume={93},
  number={12},
  pages={125113},
  year={2016},
  publisher={APS}
}

@article{dietl2008new,
  title={New magnetic field dependence of Landau levels in a graphenelike structure},
  author={Dietl, Petra and Pi{\'e}chon, Fr{\'e}d{\'e}ric and Montambaux, Gilles},
  journal={Phys. Rev. Lett.},
  volume={100},
  number={23},
  pages={236405},
  year={2008},
  publisher={American Physical Society},
  doi={10.1103/PhysRevLett.100.236405}
}

@article{carbotte2019optical,
  title={Optical properties of a semi-Dirac material},
  author={Carbotte, J. P. and Bryenton, K. R. and Nicol, E. J.},
  journal={Phys. Rev. B},
  volume={99},
  number={11},
  pages={115406},
  year={2019},
  publisher={American Physical Society},
  doi={10.1103/PhysRevB.99.115406}
}

@article{link2018out,
  title={Out-of-Bounds Hydrodynamics in Anisotropic Dirac Fluids},
  author={Link, Julia M. and Narozhny, Boris N. and Kiselev, Egor I. and Schmalian, J{\"o}rg},
  journal={Phys. Rev. Lett.},
  volume={120},
  number={19},
  pages={196801},
  year={2018},
  publisher={American Physical Society},
  doi={10.1103/PhysRevLett.120.196801}
}

@article{ye2018semidirac,
  title={Semi-Dirac point and its transport properties in two-dimensional deformed hexagonal lattice},
  author={Ye, Pei-Pei and Xu, Lei and Zhang, Jun},
  journal={Mod. Phys. Lett. B},
  volume={32},
  number={17},
  pages={1850193},
  year={2018},
  publisher={World Scientific},
  doi={10.1142/S0217984918501932}
}

@article{chan2023quantum,
  title={Quantum transport and shot noise in two-dimensional semi-Dirac system},
  author={Chan, Wei Jie and Ang, L. K. and Ang, Yee Sin},
  journal={Appl. Phys. Lett.},
  volume={122},
  number={16},
  pages={163102},
  year={2023},
  publisher={AIP Publishing},
  doi={10.1063/5.0147268}
}

@article{huang2024generation,
  title={The generation and detection of the spin-valley-polarization in semi-Dirac materials},
  author={Huang, Yupeng and Shen, R.},
  journal={Phys. Scr.},
  volume={99},
  number={2},
  pages={025927},
  year={2024},
  publisher={IOP Publishing},
  doi={10.1088/1402-4896/ad1a03}
}

@article{Saha2016Photoinduced,
title={Photoinduced Chern insulating states in semi-Dirac materials},
author={K. Saha},
journal={Phys. Rev. B},
year={2016},
volume={94},
pages={081103},
doi={10.1103/physrevb.94.081103}
}

@article{Bahamondes2024Holographic,
title={Holographic description of an anisotropic Dirac semimetal},
author={Sebastián Bahamondes and I. Salazar Landea and R. Soto-Garrido},
journal={J. High Energy Phys.},
year={2024},
volume={2024},
doi={10.1007/jhep09(2024)080}
}

@article{Liao2025Berry,
  title={Berry Curvature Dipole and Nonlinear Hall Effect in Type-II Semi-Dirac Systems},
  author={Liao, Z. and Zeng, Hui and Wang, E. and Huang, Huaqing},
  journal={Small},
  year={2025},
  pages={e2409691},
  doi={10.1002/smll.202409691}
}

@article{asafov2024collapse,
  title={Collapse of Landau levels in semi-Dirac materials},
  author={Asafov, Daniil and Pavlov, Ilia},
  journal={Phys. Rev. B},
  volume={110},
  number={12},
  pages={125126},
  year={2024},
  publisher={APS}
}

@article{Pardo2009,
  title={Half-Metallic Semi-Dirac-Point Generated by Quantum Confinement in TiO2/VO2 Nanostructures},
  author={Pardo, Victor and Pickett, Warren E.},
  journal={Phys. Rev. Lett.},
  volume={102},
  number={16},
  pages={166803},
  year={2009},
  doi={10.1103/PhysRevLett.102.166803}
}

@article{katayama2006pressure,
  title={Pressure-induced zero-gap semiconducting state in organic conductor $\alpha$-(BEDT-TTF) 2I3 salt},
  author={Katayama, Shinya and Kobayashi, Akito and Suzumura, Yoshikazu},
  journal={J. Phys. Soc. Jpn.},
  volume={75},
  number={5},
  pages={054705},
  year={2006},
  publisher={The Physical Society of Japan}
}

@article{tajima2009experimental,
  title={Experimental study of organic zero-gap conductor $\alpha$-(BEDT-TTF) 2I3},
  author={Tajima, Naoya and Kajita, Koji},
  journal={Sci. Technol. Adv. Mater.},
  volume={10},
  number={2},
  pages={024308},
  year={2009}
}

@article{monteverde2013coexistence,
  title={Coexistence of Dirac and massive carriers in $\alpha$-(BEDT-TTF) 2 I 3 under hydrostatic pressure},
  author={Monteverde, Miguel and Goerbig, MO and Auban-Senzier, Pascale and Navarin, Fabien and Henck, H and Pasquier, CR and M{\'e}zi{\`e}re, C{\'e}cile and Batail, Patrick},
  journal={Phys. Rev. B - Condens. Matter Mater. Phys.},
  volume={87},
  number={24},
  pages={245110},
  year={2013},
  publisher={APS}
}

@article{Rodin2014Strain,
  title={Strain-induced gap modification in black phosphorus},
  author={Rodin, Andrei S. and Carvalho, Alberto and Castro Neto, A. H.},
  journal={Phys. Rev. Lett.},
  volume={112},
  number={17},
  pages={176801},
  year={2014},
  publisher={American Physical Society},
  doi={10.1103/PhysRevLett.112.176801}
}

@article{Shao2024SemiDirac,
  title={Semi-Dirac Fermions in a Topological Metal},
  author={Shao, Yinming and Moon, Sangjun and Rudenko, Alexander N. and Wang, Jie and Herzog-Arbeitman, Jonah and Ozerov, M. and Graf, David and Sun, Zhiyuan and Queiroz, Raquel and Lee, Seng Huat and Zhu, Yanglin and Mao, Zhiqiang and Katsnelson, M. I. and Bernevig, B. Andrei and Smirnov, Dmitry and Millis, A. J. and Basov, D. N.},
  journal={Phys. Rev. X},
  volume={14},
  number={4},
  pages={041057},
  year={2024},
  doi={10.1103/PhysRevX.14.041057}
}

@article{Subedi2024DeepLying,
  title={Deep-lying semi-Dirac fermions in hexagonal close-packed cadmium},
  author={Subedi, A. and Behnia, K.},
  journal={J. Phys. Condens. Matter},
  volume={37},
  year={2024},
  doi={10.1088/1361-648x/ae177b}
}

@article{Zhai2011Energy,
  title={Energy spectrum of Dirac electrons on the surface of a topological insulator modulated by a spiral magnetization superlattice},
  author={Zhai, Feng and Mu, Peiyan and Chang, Kai},
  journal={Phys. Rev. B},
  volume={83},
  number={19},
  pages={195402},
  year={2011},
  publisher={American Physical Society},
  doi={10.1103/PhysRevB.83.195402}
}

@article{Li2011Anisotropic,
  title={Anisotropic surface transport in topological insulators in proximity to a helical spin density wave},
  author={Li, Qiuzi and Rossi, Enrico and Das Sarma, S.},
  journal={Phys. Rev. B},
  volume={83},
  number={8},
  pages={085110},
  year={2011},
  publisher={American Physical Society},
  doi={10.1103/PhysRevB.83.085110}
}

@article{Huang2015Chern,
  title={Emergence of a Chern-insulating state from a semi-Dirac dispersion},
  author={Huang, Huaqing and Liu, Zhirong and Zhang, Hongbin and Duan, Wenhui and Vanderbilt, David},
  journal={Phys. Rev. B},
  volume={92},
  number={16},
  pages={161115},
  year={2015},
  publisher={American Physical Society},
  doi={10.1103/PhysRevB.92.161115}
}

@article{Wu2014SemiDirac,
  title={A semi-Dirac point and an electromagnetic topological transition in a dielectric photonic crystal},
  author={Wu, Ying},
  journal={Opt. Express},
  volume={22},
  number={2},
  pages={1906--1917},
  year={2014},
  publisher={Optica Publishing Group},
  doi={10.1364/OE.22.001906}
}

@article{Milicevic2019TypeIII,
  title={Type-III and Tilted Dirac Cones Emerging from Flat Bands in Photonic Orbital Graphene},
  author={Mili{\'c}evi{\'c}, Marko and Ozawa, Tomoki and Montambaux, Gilles and Carusotto, Iacopo and Jacqmin, Timoth{\'e}e and Galopin, Erwan and Lema{\^i}tre, Aristide and Sagnes, Isabelle and Bloch, Jacqueline and Amo, Alberto},
  journal={Phys. Rev. X},
  volume={9},
  number={3},
  pages={031010},
  year={2019},
  publisher={American Physical Society},
  doi={10.1103/PhysRevX.9.031010}
}

@article{Real2020Semi-Dirac,
  title={Semi-Dirac transport and anisotropic localization in polariton honeycomb lattices},
  author={Real, B and Jamadi, O and Mili{\'c}evi{\'c}, M and Pernet, N and St-Jean, P and Ozawa, T and Montambaux, G and Sagnes, I and Lema{\^\i}tre, A and Le Gratiet, L and others},
  journal={Phys. Rev. Lett.},
  volume={125},
  number={18},
  pages={186601},
  year={2020},
  publisher={APS}
}

@article{ucho2026semidirac,
  title={Colloquium: Semi-Dirac Fermions in Quantum Matter},
  author={Uchoa, Bruno and Elsayed, Mohamed M. and Kotov, Valeri N. 
  and Shao, Yinming and Basov, Dmitri N.},
  journal={arXiv preprint arXiv:2607.20627},
  year={2026}
}

@article{hasegawa2006zero,
  title={Zero modes of tight-binding electrons on the honeycomb lattice},
  author={Hasegawa, Yasumasa and Konno, Rikio and Nakano, Hiroki and Kohmoto, Mahito},
  journal={Phys. Rev. B - Condens. Matter Mater. Phys.},
  volume={74},
  number={3},
  pages={033413},
  year={2006},
  publisher={APS}
}

@article{montambaux2009merging,
  title={Merging of Dirac points in a two-dimensional crystal},
  author={Montambaux, Gilles and Pi{\'e}chon, F and Fuchs, J-N and Goerbig, Mark O},
  journal={Phys. Rev. B - Condens. Matter Mater. Phys.},
  volume={80},
  number={15},
  pages={153412},
  year={2009},
  publisher={APS}
}

@article{montambaux2009universal,
  title={A universal Hamiltonian for motion and merging of Dirac points in a two-dimensional crystal},
  author={Montambaux, G and Pi{\'e}chon, F and Fuchs, J-N and Goerbig, MO},
  journal={Eur. Phys. J. B},
  volume={72},
  number={4},
  pages={509--520},
  year={2009},
  publisher={Springer}
}

@article{chen2018floquet,
  title={Floquet band structure of a semi-Dirac system},
  author={Chen, Qi and Du, Liang and Fiete, Gregory A.},
  journal={Phys. Rev. B},
  volume={97},
  number={3},
  pages={035422},
  year={2018},
  publisher={American Physical Society},
  doi={10.1103/PhysRevB.97.035422}
}

@article{GarcíaOlmos2024Zero,
title={Zero momentum topological insulator in 2D semi-Dirac materials},
author={Marta García Olmos and Yuriko Baba and Mario Amado and Rafael A. Molina},
journal={J. Phys. Mater.},
year={2024},
volume={7},
doi={10.1088/2515-7639/ad8378}
}

@article{GarcíaOlmos2025Spin-dependent,
title={Spin-dependent transport through edge states in 2D semi-Dirac materials with Rashba spin-orbit coupling and band inversion},
author={Marta García Olmos and Yuriko Baba and Alexander L'opez and Mario Amado and Rafael A. Molina},
journal={2D Mater.},
year={2025},
volume={12},
doi={10.1088/2053-1583/ae0f28}
}

@article{sinha2020quantum,
  title={Quantum Hall studies of a semi-Dirac nanoribbon},
  author={Sinha, Priyanka and Murakami, Shuichi and Basu, Saurabh},
  journal={Phys. Rev. B},
  volume={102},
  number={8},
  pages={085416},
  year={2020},
  publisher={APS}
}

@article{fernandez2007magnetism,
  title={Magnetism in graphene nanoislands},
  author={Fern{\'a}ndez-Rossier, Joaqu{\'\i}n and Palacios, J\_ J},
  journal={Phys. Rev. Lett.},
  volume={99},
  number={17},
  pages={177204},
  year={2007},
  publisher={APS}
}

@article{yazyev2010emergence,
  title={Emergence of magnetism in graphene materials and nanostructures},
  author={Yazyev, Oleg V},
  journal={Rep. Prog. Phys.},
  volume={73},
  number={5},
  pages={056501},
  year={2010}
}

@article{li2020uncovering,
  title={Uncovering the triplet ground state of triangular graphene nanoflakes engineered with atomic precision on a metal surface},
  author={Li, Jingcheng and Sanz, Sofia and Castro-Esteban, Jesus and Vilas-Varela, Manuel and Friedrich, Niklas and Frederiksen, Thomas and Pe{\~n}a, Diego and Pascual, Jose Ignacio},
  journal={Phys. Rev. Lett.},
  volume={124},
  number={17},
  pages={177201},
  year={2020},
  publisher={APS}
}

@article{yan2019recent,
  title={Recent advances on graphene quantum dots: from chemistry and physics to applications},
  author={Yan, Yibo and Gong, Jun and Chen, Jie and Zeng, Zhiping and Huang, Wei and Pu, Kanyi and Liu, Jiyang and Chen, Peng},
  journal={Adv. Mater.},
  volume={31},
  number={21},
  pages={1808283},
  year={2019},
  publisher={Wiley Online Library}
}

@book{gucclu2014graphene,
  title={Graphene quantum dots},
  author={G{\"u}{\c{c}}l{\"u}, Alev Devrim and Potasz, Pawel and Korkusinski, Marek and Hawrylak, Pawel and others},
  year={2014},
  publisher={Springer}
}

@article{gucclu2013zero,
  title={Zero-energy states of graphene triangular quantum dots in a magnetic field},
  author={G{\"u}{\c{c}}l{\"u}, Alev Devrim and Potasz, P and Hawrylak, P},
  journal={Phys. Rev. B - Condens. Matter Mater. Phys.},
  volume={88},
  number={15},
  pages={155429},
  year={2013},
  publisher={APS}
}

@article{potasz2012electronic,
  title={Electronic properties of gated triangular graphene quantum dots: Magnetism, correlations, and geometrical effects},
  author={Potasz, P and G{\"u}{\c{c}}l{\"u}, AD and W{\'o}js, A and Hawrylak, P},
  journal={Phys. Rev. B - Condens. Matter Mater. Phys.},
  volume={85},
  number={7},
  pages={075431},
  year={2012},
  publisher={APS}
}

@misc{landau1958quantum,
  title={Quantum Mechanics, Non-Relativistic Theory: Vol. 3 of Course of Theoretical Physics},
  author={Landau, Lev Davidovich and Lifshitz, Evgenii Mikhailovich and Sykes, John Bradbury and Bell, John Stewart and Rose, ME},
  year={1958},
  publisher={American Institute of Physics}
}

@article{li2024armchair,
  title={Armchair edge states in shear-strained graphene: Magnetic properties and quantum valley Hall edge states},
  author={Li, Ruigang and Fu, Pei-Hao and Liu, Jun-Feng and Wang, Jun},
  journal={Phys. Rev. B},
  volume={109},
  number={4},
  pages={045403},
  year={2024},
  publisher={APS}
}

@article{polat2020collapse,
  title={Collapse of the vacuum in hexagonal graphene quantum dots: A comparative study between tight-binding and mean-field Hubbard models},
  author={Polat, Mustafa and Sevin{\c{c}}li, H{\^a}ldun and G{\"u}{\c{c}}l{\"u}, Alev Devrim},
  journal={Phys. Rev. B},
  volume={101},
  number={20},
  pages={205429},
  year={2020},
  publisher={APS}
}

@article{polat2020atomic,
  title={Atomic collapse in disordered graphene quantum dots},
  author={Polat, Mustafa and G{\"u}{\c{c}}l{\"u}, Alev Devrim},
  journal={Phys. Rev. B},
  volume={102},
  number={17},
  pages={174204},
  year={2020},
  publisher={APS}
}

@phdthesis{polat2020electronic,
  title={Electronic, Magnetic and Transport Properties of Graphene Quantum Dots with Charged Impurities},
  author={Polat, Mustafa},
  year={2020},
  school={Izmir Institute of Technology (Turkey)}
}

@article{kul2020electronic,
  title={Electronic and magnetic properties of graphene quantum dots with two charged vacancies},
  author={Kul, E Bulut and Polat, Mustafa and G{\"u}{\c{c}}l{\"u}, Alev Devrim},
  journal={Solid State Commun.},
  volume={322},
  pages={114096},
  year={2020},
  publisher={Elsevier}
}

@article{dahal2010edge,
  title={Edge states in a honeycomb lattice: Effects of anisotropic hopping and mixed edges},
  author={Dahal, Hari P and Hu, Zi-Xiang and Sinitsyn, NA and Yang, Kun and Balatsky, AV},
  journal={Phys. Rev. B - Condens. Matter Mater. Phys.},
  volume={81},
  number={15},
  pages={155406},
  year={2010},
  publisher={APS}
}

@article{kul2025quantum,
  title={Quantum Monte Carlo study of artificial triangular graphene quantum dots},
  author={Kul, E Bulut and {\"O}ztarhan, G{\"o}khan and {\c{C}}{\i}nar, MN and G{\"u}{\c{c}}l{\"u}, AD},
  journal={Phys. Rev. B},
  volume={112},
  number={15},
  pages={155101},
  year={2025},
  publisher={APS}
}

@article{garcia2026corner,
  title={Corner Majorana states in semi-Dirac materials},
  author={Garc{\'\i}a-Olmos, Marta and Baba, Yuriko and Amado, Mario and Molina, Rafael A},
  journal={JPhys Mater.},
  volume={9},
  number={3},
  pages={035015},
  year={2026},
  publisher={IOP Publishing}
}

@article{xu2026floquet,
  title={Floquet second-order topological insulator in strained graphene},
  author={Xu, Yu-Wen and Wan, Xiaolin and Wang, Zi-Ming and Wang, Rui and Xu, Dong-Hui},
  journal={Phys. Rev. B},
  volume={113},
  number={20},
  pages={L201113},
  year={2026},
  publisher={APS}
}

@article{feldner2010magnetism,
  title={Magnetism of finite graphene samples: Mean-field theory compared with exact diagonalization and quantum Monte Carlo simulations},
  author={Feldner, H{\'e}lene and Meng, Zi Yang and Honecker, Andreas and Cabra, Daniel and Wessel, Stefan and Assaad, Fakher F},
  journal={Phys. Rev. B - Condens. Matter Mater. Phys.},
  volume={81},
  number={11},
  pages={115416},
  year={2010},
  publisher={APS}
}

@article{hu2019room,
  title={Room-temperature magnetism and tunable energy gaps in edge-passivated zigzag graphene quantum dots},
  author={Hu, Wei and Huang, Yi and Qin, Xinmin and Lin, Lin and Kan, Erjun and Li, Xingxing and Yang, Chao and Yang, Jinlong},
  journal={npj 2D Mater. Appl.},
  volume={3},
  number={1},
  pages={17},
  year={2019},
  publisher={Nature Publishing Group UK London}
}

@article{hasegawa2012merging,
  title={Merging Dirac points and topological phase transitions in the tight-binding model on the generalized honeycomb lattice},
  author={Hasegawa, Yasumasa and Kishigi, Keita},
  journal={Phys. Rev. B - Condens. Matter Mater. Phys.},
  volume={86},
  number={16},
  pages={165430},
  year={2012},
  publisher={APS}
}

@article{elsayed2025microscopic,
  title={Microscopic lattice model for quartic semi-Dirac fermions in two dimensions},
  author={Elsayed, Mohamed M and Kotov, Valeri N},
  journal={J. Phys. Condens. Matter},
  volume={37},
  number={31},
  pages={315501},
  year={2025},
  publisher={IOP Publishing}
}

@article{han2024zigzag,
  title={Zigzag edge ferromagnetism of triangular-graphene-quantum-dot-like system},
  author={Han, Runze and Chen, Jiazhou and Zhang, Mengyue and Gao, Jinze and Xiong, Yicheng and Pan, Yue and Ma, Tianxing},
  journal={Phys. Rev. B},
  volume={109},
  number={7},
  pages={075117},
  year={2024},
  publisher={APS}
}

@article{hagymasi2018entanglement,
  title={Entanglement and magnetism in high-spin graphene nanodisks},
  author={Hagym{\'a}si, Imre and Legeza, {\"O}rs},
  journal={Phys. Rev. B},
  volume={97},
  number={3},
  pages={035142},
  year={2018},
  publisher={APS}
}

@article{song2025janus,
  title={Janus graphene nanoribbons with localized states on a single zigzag edge},
  author={Song, Shaotang and Teng, Yu and Tang, Weichen and Xu, Zhen and He, Yuanyuan and Ruan, Jiawei and Kojima, Takahiro and Hu, Wenping and Giessibl, Franz J and Sakaguchi, Hiroshi and others},
  journal={Nature},
  volume={637},
  number={8046},
  pages={580--586},
  year={2025},
  publisher={Nature Publishing Group UK, London}
}
\end{document}